\documentclass[aps,onecolumn,prb,nofootinbib,showpacs,
preprintnumbers,amsmath,amssymb,floatfix]{revtex4-2}

\usepackage[bookmarks, linktocpage, colorlinks = true, linkcolor = blue, urlcolor  = blue, citecolor = blue, anchorcolor = green, hyperindex = true, hyperfigures]{hyperref}

\usepackage{graphicx} 
\usepackage{amssymb}
\usepackage{dcolumn}
\usepackage{multirow}
\usepackage{amsmath,amssymb}
\usepackage{slashed}
\usepackage[usenames]{color}
\usepackage{float}

\newcommand{\calI}{\mathcal{I}}
\newcommand{\bk}{\boldsymbol{k}}

\newcommand{\bp}{\boldsymbol{p}}
\newcommand{\bq}{\boldsymbol{q}}
\newcommand{\bkhat}{\hat{\boldsymbol{k}}}
\newcommand{\bphat}{\hat{\boldsymbol{p}}}
\newcommand{\bqhat}{\hat{\boldsymbol{q}}}
\newcommand{\kF}{k_{\mathrm{F}}}
\newcommand{\nuF}{N(0)}
\newcommand{\Vsinglet}{V^{\mathrm{(s)}}}
\newcommand{\Vtriplet}{V^{\mathrm{(t)}}}
\newcommand{\usinglet}{u^{\mathrm{(s)}}}
\newcommand{\utriplet}{u^{\mathrm{(t)}}}

\begin{document}

\preprint{N3AS-25-002, RIKEN-iTHEMS-Report-25}
\title{Renormalization-group approach to the Kohn-Luttinger superconductivity:\\
Amplification of the pairing gap from $\ell^4$ to $\ell$}

\author{Yuki~Fujimoto}
\email{yfujimoto@berkeley.edu}
\affiliation{Department of Physics, University of California, Berkeley, CA 94720, USA}
\affiliation{Interdisciplinary Theoretical and Mathematical Sciences Program (iTHEMS), RIKEN, Wako 351-0198, Japan}

\date{\today}

\begin{abstract}
We revisit the renormalization group (RG) analysis of the Kohn-Luttinger (KL) mechanism for superconductivity.
The KL mechanism leads to superconductivity in a system with a repulsive bare interaction.
The key ingredient is the screening effect that renders the induced interaction attractive in channels with nonzero angular momentum $\ell \neq 0$, thereby triggering the Bardeen-Cooper-Schrieffer (BCS) instability.
According to the original argument, the resulting gap is exponentially small, with its exponent scaling as $-\ell^4$.
However, the KL mechanism was originally formulated within perturbation theory, where the series is known to converge poorly in certain cases---most notably, for the p-wave paring gap induced by a repulsive s-wave contact interaction.
This poor convergence may be attributed to a divergent integrand in a specific class of diagrams containing both the BCS logarithm and the Kohn anomaly, suggesting that one must resum the Kohn anomaly contributions separately from the BCS logarithm.
In this work, we incorporate the Kohn anomaly contribution into the beta function of the RG equation governing the BCS instability near the Fermi surface.
Our solution shows that the KL gap exponent is then proportional to $-\ell$, indicating a significant enhancement of the KL mechanism beyond the previously known result.
To illustrate this, we study the spin-triplet p-wave pairing gap arising from a repulsive s-wave contact interaction and compare our RG-based results with those obtained from the Bethe-Salpeter equation in perturbation theory.
\end{abstract}

\maketitle


\section{Introduction}

The Bardeen-Cooper-Schrieffer (BCS) theory of superconductivity requires an attractive interaction for Cooper pair formation.
However, even if the bare interaction is repulsive, higher-order screening corrections can make the effective interaction attractive in channels with nonzero angular momentum $\ell \neq 0$, thus leading to the BCS instability.
This phenomenon is known as the Kohn-Luttinger (KL) mechanism~\cite{Kohn:1965zz, PhysRev.150.202} (see also Refs.~\cite{1992IJMPB...6.2471B, 2013LNP...874.....K, 2013AIPC.1550....3M} and references therein).
The corresponding pairing gap is exponentially small, with the exponent scaling as $-\ell^4$, i.e., $\ln\Delta \propto -\ell^4$.
Originally, the KL mechanism was formulated for $\ell \gg 1$, and its validity could only be extrapolated down to $\ell = 2$.
Subsequent independent studies showed that the $\ell=1$ channel also exhibits a KL-type mechanism in leading-order (LO) perturbation theory~\cite{PhysRevLett.20.187, 1988JETPL..47..614K}.

The KL mechanism has been discussed in several solid-state systems in recent years (see, e.g., Refs.~\cite{2008PhRvB..78t5431G, 2014PhRvB..89n4501N, 2014SSCom.188...61K, 2014JETP..118..995K, 2018PhRvB..98u4521L, PhysRevLett.122.026801, 2022PhRvB.105m4524Y, 2022PhRvB.105g5432C, 2024arXiv240805271H, 2024arXiv240916114Q, 2024arXiv241109664J}).
In general, the higher partial-wave ($\ell \geq 1$) pairing, which is characteristic to the KL mechanism, is known to occur in various physical systems.
A prominent example is the superfluid phase of $^3$He~\cite{Leggett:1975te, Lee:1997zzh}.
Such exotic pairing may also arise in unconventional superconductors and in cold atom experiments (see, e.g., Refs.~\cite{1994AdPhy..43..113S, BARANOV200871, 2011ARCMP...2..167K, 2012RPPh...75d2501K} and references therein).
In nuclear physics, neutron-star matter is widely believed to form a $^3P_2$ superfluid state~\cite{Hoffberg:1970vqj, Tamagaki:1970ptp, Schwenk:2003bc, Kumamoto:2024muw}, which may play a significant role in the cooling of neutron stars~\cite{Page:2010aw, Shternin:2010qi}.
Moreover, the KL mechanism has also been discussed in gauge theories in relation to color superconductivity~\cite{Schafer:2006ue}.
The application of the KL mechanism can be found in multiple physical systems across various scales since this is a generic mechanism for getting pairing from repulsion in the weak-coupling regime.

Traditionally, the KL effect is derived by solving the Bethe-Salpeter equation in perturbation theory.
The Cooper pairing gap is then extracted from a singularity in the full four-point vertex function---signaling the onset of the BCS instability---via the solution of the Bethe-Salpater equation.
Although the pairing phenomenon is inherently nonperturbative, one can still evaluate the gap in a perturbative series;
however, this series is known to converge poorly in certain cases.
For example, a numerical study~\cite{2000JETP...90..861E} of the $\ell = 1$ pairing gap from a repulsive s-wave contact interaction up to two-loop order revealed poor convergence relative to the LO results in Refs.~\cite{PhysRevLett.20.187, 1988JETPL..47..614K}.
The problematic diagrams at two-loop order include both the BCS particle-particle loop and a particle-hole loop as subdiagrams, leading to divergent integrands corresponding to the BCS logarithm and the Kohn anomaly, respectively, even though the total integral remains finite.
The Kohn anomaly is tied to the divergent integrand of the particle-hole loop diagram at momentum transfer $q = 2\kF$, where $\kF$ is the Fermi momentum.
Typically, the BCS logarithm is resummed via ladder diagrams, but the poor convergence suggests that the Kohn anomaly must also be resummed beyond the simple ladder BCS approximation.

As an alternative approach to extracting the pairing gap, one can use the renormalization group (RG) equation near the Fermi surface.
The application of the RG to the Fermi surface in higher dimensions has a relatively short history compared to the entire history of the problem;
it was intensively studied in early 1990s~\cite{BenfattoGallavotti1990,Benfatto:1990zz, Benfatto:1996ng, Feldman1990PerturbationTF, Feldman1991TheFO, Feldman1992AnIV, Feldman:1993ck, SHANKAR1991530, Shankar:1993pf, Polchinski:1992ed}, while there has been earlier applications to the Kondo problem in one dimension~\cite{1969PhRvL..23...89A, 1974JLTP...17...31N, 1975RvMP...47..773W, 1980PhRvB..21.1003K,*1980PhRvB..21.1044K}.
In nuclear physics, the RG analysis was applied to the Landau parameters of nuclear matter viewed as the Fermi liquid~\cite{Schwenk:2001hg, Schwenk:2002fq}, and it has also proved instrumental in understanding color superconductivity in the weak-coupling regime of QCD~\cite{Son:1998uk} (see also Refs.~\cite{Evans:1998ek, Evans:1998nf, Schafer:1998na}).

In this paper, we revisit the RG analysis of the KL mechanism.
Earlier RG work on the KL mechanism can be found in Refs.~\cite{Shankar:1993pf, 2001JSP...103..485S}.
An advantage of the RG method is that the aforementioned Kohn anomaly contribution emerges naturally in the RG equation through the so-called zero-sound (ZS) diagrams, which we reanalyze here (see Appendix B of Ref.~\cite{Shankar:1993pf} for the original calculation).
Although such a contribution appears as an irrelevant operator in the beta function and thus renormalizes to zero, it still affects the RG flow well before reaching the fixed point---precisely where the BCS instability sets in.
We find that incorporating the ZS diagram contribution in the beta function yields a pairing gap whose exponent scales as $-\ell$.
This differs from the standard KL expectation and indicates a substantial amplification of the KL pairing gap.

The paper is organized as follows.
In Sec.~\ref{sec:KL}, we review the conventional argument for the KL mechanism.
In Sec.~\ref{sec:RG}, we present the RG approach near the Fermi surface, discussing the formation of the BCS superconducting state in Sec.~\ref{sec:RGBCS} and how the KL mechanism appears through the ZS diagram in Sec.~\ref{sec:RGKL}.
The readers who are familiar with the KL mechanism can skip Sec.~\ref{sec:KL}, and those familiar with the RG approach can skip Secs.~\ref{sec:RGsetup} and \ref{sec:RGBCS}.
We then compute the beta function from the ZS diagram in Sec.~\ref{sec:betaZS}.
In Sec.~\ref{sec:enhance}, we show how the inclusion of the ZS contribution in the RG flow amplifies the KL mechanism;
this is our main result.
We exemplify such an enhancement by the p-wave pairing gap from the bare repulsive s-wave contact interaction and compare the RG approach with the conventional perturbation theory argument in Sec.~\ref{sec:tripletpwave}.
Finally, we summarize the conclusions and discuss future outlook.
An appendix details the extension of the RG equation to include spin degrees of freedom.

\section{Kohn-Luttinger Mechanism}
\label{sec:KL}

\begin{figure}
    \centering
    \includegraphics[width=0.4\columnwidth]{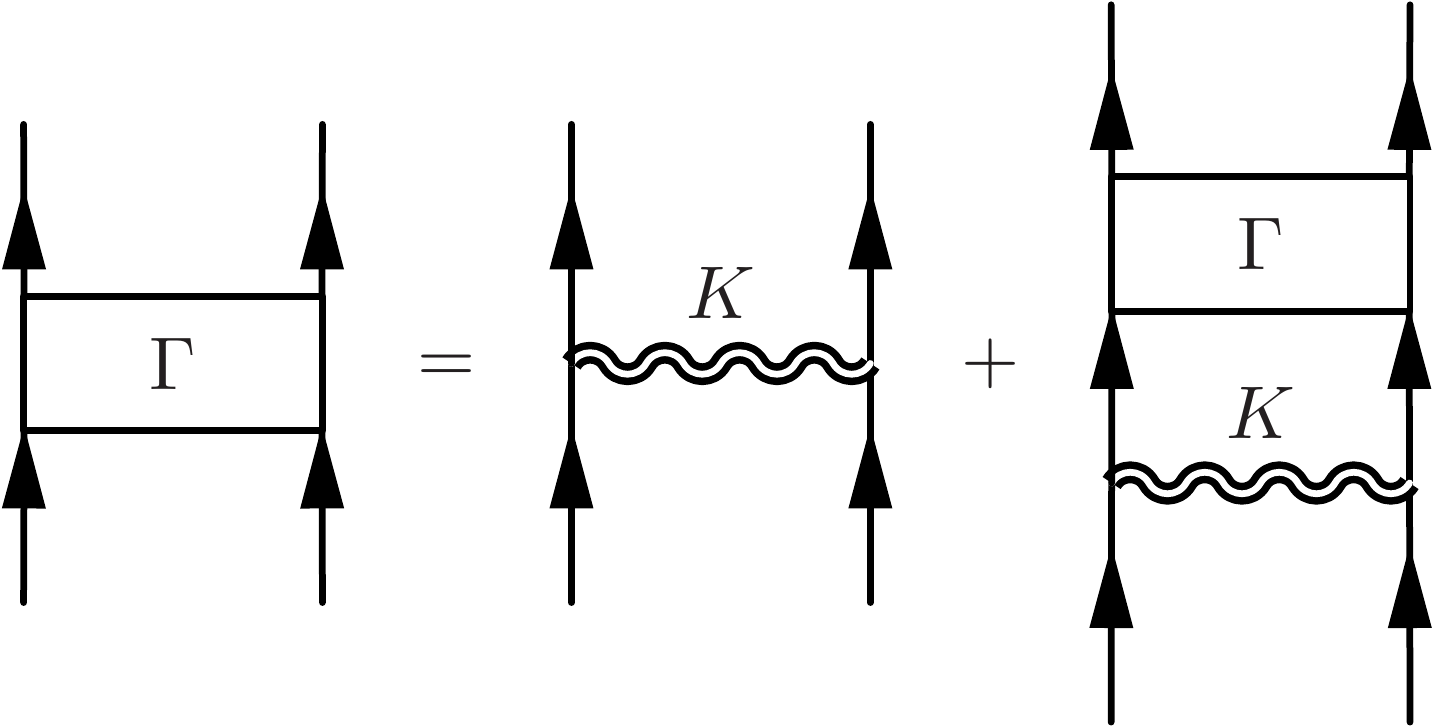}\\
    \includegraphics[width=0.8\columnwidth]{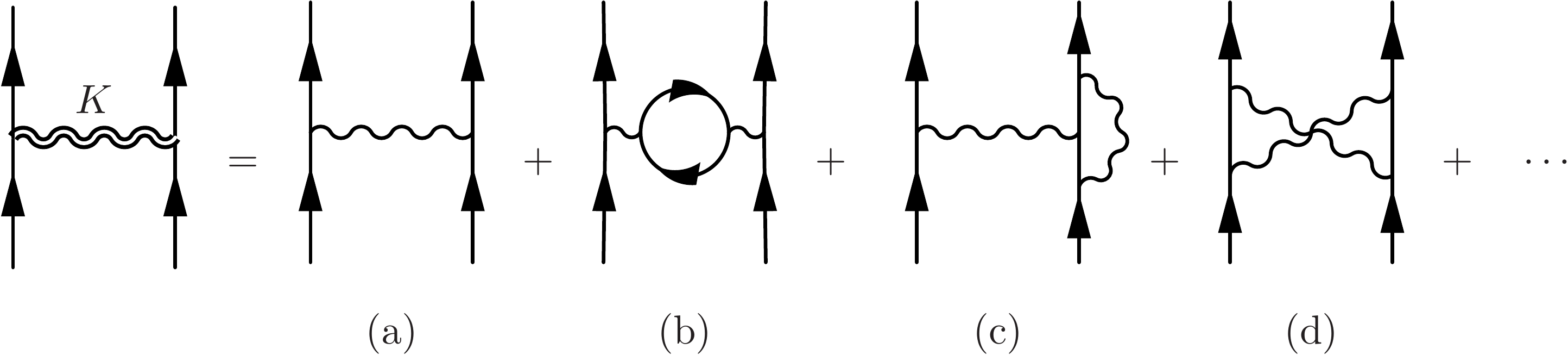}
    \caption{Top panel: The diagrammatic representation of the Bethe-Salpeter equation. The solution to this equation gives the proper vertex $\Gamma$ from which one can read out the BCS instability.
    Bottom panel: The Kohn-Luttinger diagrams of $t$-channel particle-particle interactions up to one-loop order that contribute to the irreducible interaction $K$, which is represented by the double wiggly line.
    }
    \label{fig:KL}
\end{figure}

Here, we review the KL mechanism.
The conventional argument by Kohn and Luttinger in Ref.~\cite{Kohn:1965zz} relies on a one-loop perturbative analysis.
They considered the scattering of a particle pair with momenta $(\bk_1, -\bk_1) \to (\bk_3, -\bk_3)$, described by a weak, short-range BCS potential $U(\theta) = U(\bk_1, \bk_3)$, where the scattering angle $\theta$ is defined via $\cos \theta = \bkhat_1 \cdot \bkhat_3$.
The emergence of superconducting/superfluid pairing is associated with a singularity in the two-particle vertex function $\Gamma$ in the particle-particle (BCS) channel ar zero total momentum and frequency.
This proper vertex $\Gamma$ is a solution of the Bethe-Salpeter equation shown in the top panel of Fig.~\ref{fig:KL}, where the double wiggly lines represent the irreducible interaction $K$ in the BCS channel.
Kohn and Luttinger then analyzed $K$ for large $\ell$ up to one-loop order.
It consists of the tree-level potential $U(\theta)$ plus the $O(U^2(\theta))$ terms arising from screening at $\theta = \pi$, as depicted in the bottom panel of Fig.~\ref{fig:KL}.
They argued that the non-analytic scattering near $\theta=\pi$ contributes $\sim 1/\ell^4$ to $K_{\ell}$, where as the analytic tree-level contribution $U(\theta)$ behaves like $e^{-\ell}$ for large $\ell$.
Consequently, $K_{\ell}$ can become attractive, particularly for odd $\ell$.

The analytic understanding is as follows.
The diagram (a) corresponds to the part of the irreducible interaction
\begin{equation}
    K^{\mathrm{(a)}}(\bk_1, \bk_3) = U(\bk_1, \bk_3)\,.
\end{equation}
The diagram (b) corresponds to
\begin{align}
    K^{\mathrm{(b)}}(\bk_1, \bk_3)
    = 2\int \frac{d^4 P}{(2\pi)^4} \frac{U^2(\bk_1, \bk_3)}{(i \omega - \epsilon_{\bp})(i\omega - \epsilon_{\bp+\bq})} \,,
\end{align}
where $\bq = \bk_1 - \bk_3$ is the momentum transfer, $P^\mu = (\omega, \bp)$ is the Euclidean four vector, and $\epsilon_{\bk} = \bk^2 / (2m) - \mu$ is the energy relative to the Fermi surface with $\mu$ being the chemical potential.
After performing the $\omega$-integral, one finds
\begin{align}
    K^{\mathrm{(b)}}(\bk_1, \bk_3)
    = 2\int \frac{d^3 \bp}{(2\pi)^3} U^2(\bk_1, \bk_3) \frac{\Theta(|\bp + \bq| - \kF) - \Theta(|\bp| - \kF)}{\epsilon_{\bp} - \epsilon_{\bp + \bq}}\,.
\end{align}
The integrand is singular at $\bq = - 2\bp$, which is commonly referred to as the Kohn anomaly~\cite{Kohn:1959zz}, so the entire integration is dominated by the contribution at $\bq \simeq -2\kF$.
Therefore, we approximate that the scattering amplitudes in the above expression with $U(\pi)$, which is the scattering amplitude of the back-to-back scattering at $\theta = \pi$ in the BCS channel.
We also note that the integrand is also singular at $\bq = 0$, but the step function reduces to the delta function in this limit, so this point does not contribute to the integral.
Therefore, $K^{\mathrm{(b)}}$ can be approximated as
\begin{align}
K^{\mathrm{(b)}}(\bk_1, \bk_3)
    \simeq 2 U^2(\pi) \Pi(\omega = 0, q)\,,
\end{align}
where the Lindhard function $\Pi(\omega=0, q)$ is defined as~\cite{1971qtmp.book.....F}
\begin{align}
    \Pi(\omega = 0, q)
    &\equiv \int \frac{d^3 \bp}{(2\pi)^3} \frac{\Theta(|\bp + \bq| - \kF) - \Theta(|\bp| - \kF)}{\epsilon_{\bp} - \epsilon_{\bp + \bq}} \,,\notag\\
    &= \frac{\nuF}{2} \left[-1 + \frac{\kF}{q} \left(1 - \frac{q^2}{4\kF^2}\right) \ln \left(\frac{2\kF - q}{2\kF + q}\right) \right]\,,
    \label{eq:lindhard}
\end{align}
where $\nuF = m \kF / (2 \pi^2)$ is the density of states on the Fermi surface.
Since the integral is dominated by the Kohn anomaly at $q = 2\kF$, i.e. $\theta = \pi$, one can only take out the part of the Lindhard function corresponding to the Kohn anomaly: $\hat{\Pi}(q) \propto \nuF (2\kF - q) \ln(2\kF - q)$.
Note that although the integrand is singular, the integrated Lindhard function is regular.
By substituting $q = \sqrt{2(1-\cos\theta)}\kF$ in the formula above, one obtains
\begin{equation}
    \hat{\Pi}(\cos\theta) = \frac{\nuF}{8}(1 + \cos\theta) \ln (1 + \cos\theta)\,.
\end{equation}
The part of $K^{\mathrm{(b)}}$ corresponding to the Kohn anomaly is then
\begin{align}
    \hat{K}^{\mathrm{(b)}}(\cos\theta)
    = 2 U^2(\pi) \hat{\Pi}(\cos\theta)\,.
    \label{eq:Khatb}
\end{align}
Likewise, the part of the irreducible interactions corresponding to the Kohn anomaly from the diagrams (c) and (d) in Fig.~\ref{fig:KL} are
\begin{align}
    \hat{K}^{\mathrm{(c)}}(\cos\theta)
    &= -2 U(0) U(\pi) \hat{\Pi}(\cos\theta)\,,\label{eq:Khatc}\\
    \hat{K}^{\mathrm{(d)}}(\cos\theta)
    &= - U^2(0) \hat{\Pi}(-\cos\theta)\,.
    \label{eq:Khatd}
\end{align}
Then, the part of the total irreducible interaction that corresponds to the Kohn anomaly $\hat{K}(\cos\theta)$ is given by the sum of Eqs.~(\ref{eq:Khatb}--\ref{eq:Khatd}):
\begin{align}
    \hat{K}(\cos\theta) = \hat{K}^{\mathrm{(b)}}(\cos\theta) + \hat{K}^{\mathrm{(c)}}(\cos\theta) + \hat{K}^{\mathrm{(d)}}(\cos\theta)\,.
\end{align}

Now, we perform the partial wave expansion
\begin{equation}
    \hat{K}(\cos\theta) = \sum_{\ell = 0}^{\infty} (2\ell + 1) \hat{K}_{\ell} P_{\ell}(\cos\theta)\,,
    \label{eq:partialwave}
\end{equation}
or inversely, $\hat{K}_{\ell} = \frac12 \int_{-1}^{1} d\cos\theta \, \hat{K}(\cos\theta) P_{\ell}(\cos\theta)$.
One can obtain $\hat{K}_{\ell}$ for $\ell \geq 2$ by using the Rodrigues's formula,
\begin{equation}
    P_{\ell}(x) = \frac{1}{2^{\ell}\ell!} \frac{d^{\ell}}{dx^{\ell}} \left[(x^2 - 1)^\ell\right]\,,
\end{equation}
and integrating by parts $\ell$ times:
\begin{equation}
    \hat{K}_{\ell} = - \frac{\nuF}{4}\left\{U^2(0) + 2(-1)^{\ell} \left[U(0) U(\pi) - U^2(\pi) \right]\right\} \frac{1}{(\ell - 1) \ell (\ell + 1) (\ell + 2)}\,.
\end{equation}
At large $\ell$, $\hat{K}_{\ell}$ scales as $\sim 1/ \ell^4$.
In particular, for $\ell$ odd, this is always negative:
\begin{equation}
    \hat{K}_{\ell} \simeq - \frac{\nuF}{4}\left\{\left[U(0) - U(\pi) \right]^2 + U^2(\pi)\right\} \frac{1}{\ell^4} < 0\,,
\end{equation}
and thus the interaction becomes attractive.
For constant $U(\theta) = U$ and for any $\ell \geq 2$, $\hat{K}_{\ell}$ is
\begin{equation}
    \hat{K}_{\ell} = - \frac{\nuF U^2}{4 \ell^4} < 0\,.
\end{equation}
For $\ell \geq 2$, one can approximate $K_{\ell} \simeq \hat{K}_{\ell}$.
For $\ell = 1$, a separate argument from the above consideration is necessary~\cite{PhysRevLett.20.187, 1988JETPL..47..614K}.
It turns out that the attractive interaction survives even down to $\ell = 1$, and for constant $U$, one finds the exact $K_{\ell = 1}$:
\begin{equation}
    K_{\ell = 1} = - \nuF U^2 \frac{2\ln 2 - 1}{5} \,.
\end{equation}

The pairing gap $\Delta$ can be read out from the solution of the Bethe-Salpeter equation above, and one finds
\begin{equation}
    \Delta \sim \mu \exp\left(\frac{1}{\nuF K_{\ell}}\right)\,.
\end{equation}
Therefore, the index is proportional to $\ell^4$, i.e., $\ln(\Delta / \mu) \propto -\ell^4$.

\section{BCS instability from the renormalization group}
\label{sec:RG}

In this section, we review the basic notion of the RG analysis of the interacting fermion near the Fermi surface.
One can apply this method to see the formation of the BCS superconducting state.

\subsection{Renormalization group near the Fermi surface}
\label{sec:RGsetup}

We consider the effective theory with the kinetic term of the action as
\begin{align}
    S_0 = \int_{||\bk| - \kF| < \Lambda}\frac{d^4 K}{(2\pi)^4} \bar{\psi}(K) (i\omega - \epsilon_{\bk}) \psi(K)\,,
\end{align}
where $K^\mu = (\omega, \bk)$ is the Euclidean four vector.
We restrict the effective theory to contain only the fermions located in a thin shell surrounding the Fermi surface with a cutoff $\Lambda$, i.e., $\big||\bk| - \kF \big| < \Lambda$.
As the cutoff scales to zero, the momentum must scale toward the Fermi surface.
We decompose the momentum as
\begin{equation}
    \bk = (\kF + l) \bkhat\,.
\end{equation}
Then the energy relative to the Fermi surface is expressed as
\begin{equation}
    \epsilon_{\bk} = \frac{\bk^2}{2m} - \frac{\kF^2}{2m} \simeq v l + O\left(\frac{l^2}{\kF^2}\right)\,,
\end{equation}
where $v = \kF / m$ is the Fermi velocity.
Each fermion state near the Fermi surface is specified by $(\omega, l, \bkhat)$.
As we saw above, the fermion mode in the effective theory is limited to $|l| < \Lambda$.

\begin{figure}
    \centering
    \includegraphics[width=0.8\columnwidth]{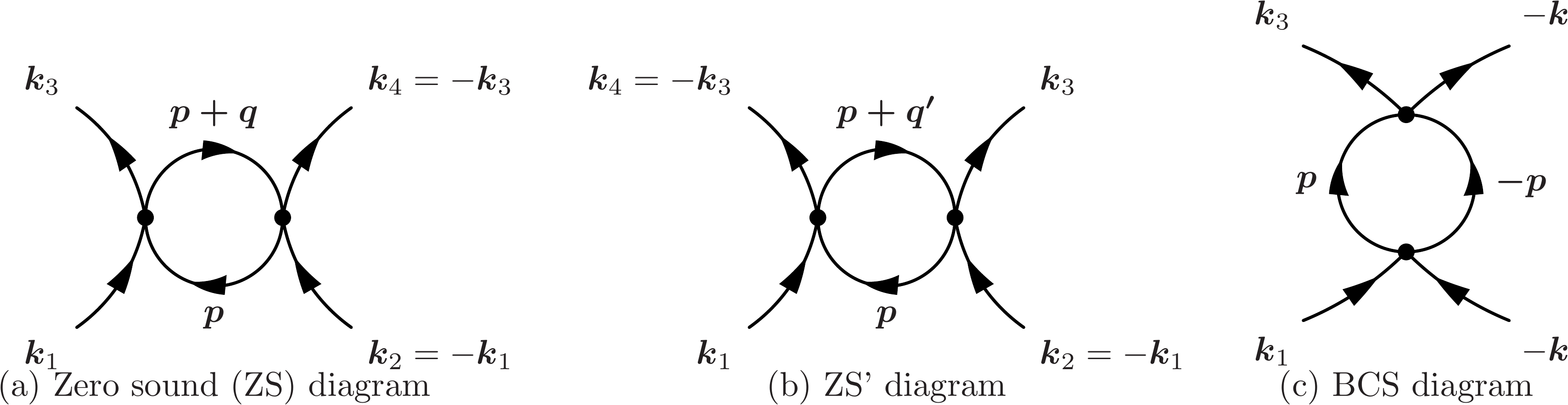}
    \caption{The skeleton one-loop diagrams that renormalize the quartic coupling. We write the loop momentum as $\bp$ and the momentum transfer as $\bq = \bk_1 - \bk_3$ and $\bq' = \bk_1 + \bk_3$, respectively.}
    \label{fig:ZS}
\end{figure}

Let us apply the RG transformation to the quartic coupling of the following form:
\begin{align}
    \label{eq:quartic}
    \delta S
    = \frac{1}{4} \left(\prod_{i=1}^4 \int_{|l| < \Lambda} \frac{d^4 K_i}{(2\pi)^4} \right)\bar{\psi}_{\sigma_4}(l_4) \bar{\psi}_{\sigma_3}(l_3)  \psi_{\sigma_2}(l_2) \psi_{\sigma_1}(l_1) u_{\sigma_1 \sigma_2 \sigma_3 \sigma_4}(l_1, l_2, l_3, l_4) (2\pi)^4 \delta^{(4)}(l_1 + l_2 - l_3 - l_4) \,,
\end{align}
where $\sigma_i$ is the spin of the $i$-th fermion.
Below, we suppress the spin indices and consider spinless fermions for simplicity.
We define the following shorthand notations for $\psi$, in which we abbreviate $\omega_i$ and $\bkhat_i$, and for the delta function:
\begin{align}
    \psi(l_i) &\equiv \psi(\omega_i, l_i, \bkhat_i)\,,\\
    \delta^{(4)}(l_1 + l_2 - l_3 - l_4) & \equiv
    \delta(\omega_1 + \omega_2 - \omega_3 - \omega_4)\delta^{(3)}[\kF(\bkhat_1 + \bkhat_2 - \bkhat_3 - \bkhat_4) + l_1 \bkhat_1 + l_2\bkhat_2 - l_3 \bkhat_3 - l_4 \bkhat_4]\,.
    \label{eq:deltal}
\end{align}
The RG procedure works as follows.
The cutoff is reduced from $\Lambda$ to $\Lambda/s$ with $s>1$ so that all fermion states with fast modes $\Lambda/s \leq |l| < \Lambda$ are integrated out.
After the renormalized effective action $S'$, which is integrated over only the slow modes $0\leq |l| < \Lambda/s$, is obtained, we express the renormalized effective action in terms of rescaled momenta
\begin{equation}
    \omega' = s \omega\,,\qquad
    l' = s l\,,\qquad
    \bkhat' = \bkhat\,,
\end{equation}
which now go all the way to $\Lambda$ again, and in terms of the rescaled field
\begin{equation}
    \psi'(l') = s^{-3/2} \psi(l'/s)\,.
\end{equation}
The scaling of the fermion field $s^{-3/2}$ is chosen so that the renormalized noninteracting action $S_0'$ has the same coefficient as $S_0$ before the mode elimination.
Since the quartic coupling mixes the slow and fast modes, the renormalized quartic coupling is
\begin{align}
    \delta S' &= \langle \delta S \rangle_{0>} + \frac12 \left( \langle \delta S^2 \rangle_{0>} - \langle \delta S \rangle_{0>}^2 \right) + \cdots\,,\notag \\
    &\equiv \delta S'_{\mathrm{tree}} + (\delta S'_{\mathrm{ZS}} + \delta S'_{\mathrm{ZS'}} + \delta S'_{\mathrm{BCS}}) + \cdots \,.
    \label{eq:deltaS}
\end{align}
where $\langle \cdot \rangle_{0>}$ refers to the averages with respect to the fast modes with action $S_0$.
The first term corresponds to the renormalization of the coupling at the tree-level and the second term in the bracket corresponds to one-loop corrections in perturbation theory expressed by the Feynman diagrams shown in Fig.~\ref{fig:ZS}.

The tree-level action is renormalized as
\begin{align}
    \delta S'_{\mathrm{tree}}
    &= \frac{1}{4} \left(\prod_{i=1}^4 \int_{|l| < \Lambda / s} \frac{d^4 K_i}{(2\pi)^4} \right)\bar{\psi}(l_4) \bar{\psi}(l_3) \psi(l_2) \psi(l_1) u(l_1, l_2, l_3, l_4) (2\pi)^4 \delta^{(4)}(l_1 + l_2 - l_3 - l_4) \,, \notag \\
    &= \frac{1}{4} s^{-2} \left(\prod_{i=1}^4 \int_{|l'| < \Lambda} \frac{d^4 K_i'}{(2\pi)^4} \right)\bar{\psi}' (l_4') \bar{\psi}'(l_3') \psi'(l_2') \psi'(l_1')  u(\tfrac{l_1'}{s}, \tfrac{l_2'}{s}, \tfrac{l_3'}{s}, \tfrac{l_4'}{s}) (2\pi)^4 s^2 \delta^{(4)}(l_1' + l_2' - l_3' - l_4') \,.
\end{align}
In the last line, in order for the delta function $\delta^{(4)}(\tfrac{l_1'}{s} + \tfrac{l_2'}{s} - \tfrac{l_3'}{s} - \tfrac{l_4'}{s})$ to scale as $s^2\delta^{(4)}(l_1' + l_2' - l_3' - l_4')$, the directions of the momenta have to fulfill the condition
\begin{equation}
    \bkhat_1 + \bkhat_2 - \bkhat_3 - \bkhat_4 = 0\,.
    \label{eq:kinematic}
\end{equation}
This is because when $\bkhat_1 + \bkhat_2 - \bkhat_3 - \bkhat_4 \neq 0$, the argument of the second delta function in Eq.~\eqref{eq:deltal} does not scale as $s$;  the first term is much larger compared to the latter term as $\kF \gg l_i$.

At the same time, $\delta S'$ can be written in terms of the rescaled momenta:
\begin{equation}
    \delta S' = \frac{1}{4} \left(\prod_{i=1}^4 \int_{|l'| < \Lambda} \frac{d^4 K_i'}{(2\pi)^4} \right)\bar{\psi}' (l_4') \bar{\psi}'(l_3') \psi'(l_2') \psi'(l_1')  u'(l_1', l_2', l_3', l_4') (2\pi)^4 \delta^{(4)}(l_1' + l_2' - l_3' - l_4') \,.
\end{equation}
From Eq.~\eqref{eq:deltaS}, the renormalized coupling $u'$ is obtained:
\begin{align}
    u'(l_1', l_2', l_3', l_4') = u(l_1' / s, l_2' / s, l_3' / s, l_4' / s) + \delta u_{\mathrm{ZS}} + \delta u_{\mathrm{ZS'}} + \delta u_{\mathrm{BCS}} + \cdots\,,
\end{align}
where $\delta u_{\mathrm{ZS, ZS', BCS}}$ corresponds to the ZS, ZS', and BCS diagrams, respectively.
From this, one finds the RG equation
\begin{align}
    \frac{d u}{dt}
    &= \left. \frac{d u_{\mathrm{tree}}}{dt}\right|_{s=1} + \left. \frac{d \delta u_{\mathrm{ZS}}}{dt}\right|_{s=1} + \left. \frac{d \delta u_{\mathrm{ZS'}}}{dt}\right|_{s=1} + \left. \frac{d \delta u_{\mathrm{BCS}}}{dt}\right|_{s=1} + \cdots\,, \notag\\
    &\equiv \beta_{\mathrm{tree}} + \beta_{\mathrm{ZS}} + \beta_{\mathrm{ZS'}} + \beta_{\mathrm{BCS}}+ \cdots \,.
    \label{eq:RGeq}
\end{align}
where we defined the scale parameter $t \equiv \ln s$\,.
Usually, $\beta_{\mathrm{tree}} = 0$.
The one-loop beta functions $\beta_{\mathrm{ZS}}$ and $\beta_{\mathrm{ZS'}}$ on the right hand side turn out to be irrelevant operator except for $\beta_{\mathrm{BCS}}$.
Also, the subleading contributions to beta functions in the loop expansion are also irrelevant.

\subsection{Renormalization group equation in the BCS kinematics}
\label{sec:RGBCS}

The condition~\eqref{eq:kinematic} constrains the possible kinematics in order for the coupling to be marginal.
Here, we limit ourselves to the BCS kinematics with $\bkhat_1 = -\bkhat_2$ and $\bkhat_3 = -\bkhat_4$, and the coupling function essential for BCS instability is characterized by the one in the BCS channel:
\begin{equation}
    V(\theta) \equiv V(\bk_1, \bk_3) \equiv u(\bkhat_1, -\bkhat_1, \bkhat_3, -\bkhat_3)\,,
    \label{eq:Vdef}
\end{equation}
where $\bkhat_1 \cdot \bkhat_3 = \cos \theta$.
The correction to $V(\theta)$ from the BCS diagram is
\begin{align}
    \delta V_{\mathrm{BCS}}(\bk_1, \bk_3)
    &= - \frac12 \int_{\Lambda /s \leq |l_{\bp}| < \Lambda}\frac{d^4 P}{(2\pi)^4} 
    \frac{V(\bk_1, \bp) V(\bp, \bk_3)}{(i \omega - \epsilon_{\bp})(-i\omega - \epsilon_{-\bp})}\,,
\end{align}
where $l_{\bp} = |\bp| - \kF$ and we set $\epsilon_{\bk_1} = 0$.
The beta function associated with this BCS diagram is
\begin{align}
    \beta_{\mathrm{BCS}}&= \left. \frac{d \delta V_{\mathrm{BCS}}}{dt} \right|_{s=1} = - \frac{\nuF}{2} \int \frac{d \bphat}{4\pi} V(\bk_1, \bp) V(\bp, \bk_3) \,.
\end{align}
In the partial-wave expansion as in Eq.~\eqref{eq:partialwave}, the BCS beta function becomes
\begin{equation}
    \beta_{\mathrm{BCS}, \ell} = - \frac{\nuF}{2} V_{\ell}^2\,. 
\end{equation}
Hereafter, the subscript $\ell$ refers to the quantity in the partial wave channel $\ell$.
Also, we understand that the coupling function $V(\theta; t)$ or $V_{\ell}(t)$ has $t$ dependence, however we sometimes abbreviate it in the argument of the function for the sake of simplicity.
For the moment, let us ignore the tree-level beta function $\beta_{\mathrm{tree}}$ and the irrelevant operators $\beta_{\mathrm{ZS}}$ and $\beta_{\mathrm{ZS'}}$ in the RG equation~\eqref{eq:RGeq}.
It becomes
\begin{align}
    \frac{d V_{\ell}(t)}{dt} = \beta_{\mathrm{BCS},\ell}\,.
\end{align}
A solution to this RG equation is
\begin{equation}
    V_{\ell}(t) = \frac{V_{\ell}(0)}{1 + \nuF V_{\ell}(0) t / 2}\,.
    \label{eq:RGsolution}
\end{equation}
If the interaction is repulsive in all channels at $t = 0$, i.e., all $V_{\ell}$ is positive, the quartic coupling renormalizes to zero at the Fermi surface, $t = \infty$.
However, if there is an attractive channel, i.e., one of $V_{\ell}(0)$ is negative, it will reach a pole at $t = - 2 / [\nuF V_{\ell}(0)]$, which exhibits the BCS instability of the Fermi surface for any attractive interaction.
At the BCS singularity, the typical energy scale $\Lambda$ can be identified as the BCS pairing gap, $\Lambda \simeq \Delta$.
The scale parameter can be expressed as $t = - \ln (\Lambda / \mu)$, and thus the pairing gap can be read out as
\begin{equation}
    \Delta \simeq \mu \exp\left(\frac{2}{\nuF V_{\ell}(0)}\right)\,.
\end{equation}
The channel having the largest negative $V_{\ell}(0)$ reaches the pole first, and hence develops the pairing gap.
Solving the RG equation can be regarded as the summation of the BCS diagrams in ladder shown in Fig.~\ref{fig:ZS}~(c).
The effect is common to the ladder summation through the Bethe-Salpeter equation shown in the top panel of Fig.~\ref{fig:KL}.

\subsection{Kohn-Luttinger mechanism from the renormalization group}
\label{sec:RGKL}

We now restate the KL effect in the language of the RG framework, following Refs.~\cite{Shankar:1993pf, 2001JSP...103..485S}.
In general, the condition for the occurence of the BCS instability (i.e., the RG flow hitting the singularity) is $V_{\ell}(t=0) < 0$, as one can clearly see from Eq.~\eqref{eq:RGsolution}.
If the interaction is initially attractive, $V_{\ell}$ becomes negative under the RG flow, and the BCS instability is thus guaranteed.
By contrast, if the bare interaction is repulsive, the tree-level value $V_{\ell}(t=0)$ is positive, and no BCS instability occurs at that order.
However, according to the KL mechanism, this conclusion can be altered.
Specifically, two factors can drive the system toward an instability even when the bare interaction is repulsive:
(1) a modification of the initial condition $V_{\ell}(t=0)$ due to loop corrections, and
(2) a modification of the RG equation itself.

Regarding the modification to the initial condition for the RG flow, this is precisely what Kohn and Luttinger originally identified in Refs.~\cite{Kohn:1965zz, PhysRev.150.202}.
They noted that $V_{\ell}(t=0)$ can become negative in the large-$\ell$ limit, even for a bare repulsive interaction.
Specifically, for some short-ranged interactions, the tree-level amplitude scales as $V_{\ell}(0) \sim e^{-\ell}$ for $\ell \gg 1$.
When multiplied by the Legendre polynomials $P_{\ell}(\cos\theta)$ and summed (including derivatives with respect to $\cos \theta$), these terms converge to a fixed analytic function $V(\theta)$.
Meanwhile, the one-loop corrections computed in perturbation theory behaves as $V_{\ell}(0) \sim -1 / \ell^4 $, arising from the singularity at $\theta = 0$ and $\pi$ in the particle-hole channel, as reviewed in Sec.~\ref{sec:KL}.
Hence, in large-$\ell$ channels, the induced attraction from the one-loop corrections overwhelms the tree-level repulsion, making $V_{\ell}(0)$ negative overall.

Turning to the modification of the RG equation, it is tied to an irrelevant operator that cannot be ignored.
Shankar first highlighted this point in his seminal review~\cite{Shankar:1993pf}, and it was also discussed in Ref.~\cite{2001JSP...103..485S}.
Up to one-loop order, including the contribution of the irrelevant operator, the full RG equation reads
\begin{align}
    \label{eq:RGeqVl}
    \frac{d V_{\ell}(t)}{dt} = \beta_{\mathrm{ZS},\ell} + \beta_{\mathrm{ZS'},\ell} + \beta_{\mathrm{BCS},\ell}\,.
\end{align}
The beta functions $\beta_{\mathrm{ZS},\ell}$ and $\beta_{\mathrm{ZS'},\ell}$ scale as $-e^{-t}$ and $(-1)^{\ell} e^{-t}$, respectively, and will be calculated in Sec.~\ref{sec:betaZS}.
Although these irrelevant operators renormalize to zero at large $t$, they cannot simply be set to zero from the start without affecting the intermediate RG flow.
They do not alter the final fixed point;
however, they do modify how the system evolves before that point is reached.
Since the BCS instability arises (i.e., the RG flow hits a pole) \emph{before} the flow reaches the fixed point, these negative contributions from the ZS and ZS' diagrams shift the flow and thus change where the BCS singularity occurs.
As a result, the gap parameter exhibits a different dependence on the angular momentum $\ell$, which we will explain in Sec.~\ref{sec:enhance}.

Since both the initial-condition modification and the RG-equation modification stem from the particle-hole diagram, one might suspect double counting.
However, these two modifications represent distinct contributions, so no double counting occurs.
The reasoning is as follows.
Consider the coupling function $u$ in Eq.~\eqref{eq:quartic} to be expanded perturbatively in terms of a small parameter $\lambda$ in the underlying field theory, i.e.,
\begin{equation}
    u = u_1 \lambda + u_2 \lambda^2\,,
\end{equation}
where $u_1$ and $u_2$ are certain functions.
The function $u$ need not be a tree-level amplitude in the underlying field theory.
For example, in Ref.~\cite{Son:1998uk}, a resummed propagator was used instead of the bare one.
On the left hand side of the RG equation~\eqref{eq:RGeqVl}, we then have
\begin{equation}
    \frac{du}{dt} = \frac{d}{dt} (u_1 \lambda + u_2 \lambda^2)\,.   
\end{equation}
Meanwhile, because the beta function is proportional to $u^2$, the beta function on the right-hand side of the RG equation behaves as 
\begin{equation}
    \beta \propto u^2 = (u_1 \lambda + u_2 \lambda^2)^2 = u_1^2 \lambda^2 + O(\lambda^3)\,.
\end{equation}
In our discussion, we retain terms up to $O(\lambda^2)$ on both sides of the RG equation.
Then, on the one hand, the modification to the initial condition $V_{\ell}(0)$ corresponds to the term $d(u_2 \lambda^2)/dt$ on the left-hand side, with $u_2 \sim 1/\ell^4$.
On the other hand, the modifications  $\beta_{\mathrm{ZS},\ell}$ and $\beta_{\mathrm{ZS'},\ell}$ to the RG equation arise from the term proportional to $u_1^2 \lambda^2$ on the right-hand side.
Because this RG modification stems from $u_1$, rather than $u_2$, it is independent of the contribution labeled by $u_2$ from the particle-hole diagram.
Later, in Eq.~\eqref{eq:gammaRG}, we will verify \emph{a posteriori} that keeping contributions up to $O(\lambda^2)$ in the RG equation is consistent, in the sense that it reproduces the lowest-order result for the gap parameter in a theory with bare s-wave contact repulsion.

\section{Computation of the beta function from the zero sound diagram}
\label{sec:betaZS}

We now evaluate the ZS diagram contribution $\delta V_{\mathrm{ZS}}$ to the beta function.
This term is usually neglected because it is an irrelevant operator that renormalizes to zero, and thus does not affect the RG flow toward the fixed point.
However, as noted above, we are interested in pairing near the Fermi surface, where the irrelevant component can still affect the flow away from the fixed point and hence shift the location of the BCS singularity.

In Ref.~\cite{Shankar:1993pf}, Appendix B, this calculation was carried out with a soft momentum cutoff.
Here, we perform the same calculation using a hard cutoff and reanalyze the integral.

There are two ways to define the beta function: the modern Kadanoff-Wilson approach and the field-theoretic approach, following the classification in Ref.~\cite{Shankar:1993pf}.
At one-loop order, both definitions yield the same expression.
We used the former approach above.
Now, we employ the latter approach to compute the ZS diagram.
In this field-theoretic framework, the coupling function is defined via the scattering amplitude with a UV cutoff in the Feynman integrals;
the beta function is then obtained by imposing cutoff independence of the full scattering amplitude.
At one-loop order, the full scattering amplitude $\Gamma(\theta)$ is
\begin{equation}
    \Gamma(\theta) = V(\theta) + V_{\mathrm{ZS}} + V_{\mathrm{ZS'}} + V_{\mathrm{BCS}}\,.
\end{equation}
Since $\Gamma$ is independent of the cutoff $\Lambda = \mu e^{-t}$, we have $d \Gamma / dt = 0$.
Hence, the beta function for $V(\theta)$ is found to be
\begin{align}
    \beta[V(\theta)] \equiv \frac{d V(\theta)}{dt} 
    &= - \frac{d V_{\mathrm{ZS}}}{dt} - \frac{d V_{\mathrm{ZS'}}}{dt}  - \frac{d V_{\mathrm{BCS}}}{dt}\,, \notag \\
    &\equiv \beta_{\mathrm{ZS}} + \beta_{\mathrm{ZS'}}+ \beta_{\mathrm{BCS}}\,.
    \label{eq:betaV}
\end{align}

The loop correction to $V$ from the ZS diagram is
\begin{align}
  V_{\mathrm{ZS}}(\bk_1, \bk_3)
  =  \int_{|l_{\bp}|, |l_{\bp + \bq}| < \Lambda} \frac{d^4 P}{(2\pi)^4}
    \frac{u(\bk_1, \bp, \bk_3, \bp + \bq) u(-\bp, -\bk_1, -(\bp+\bq), -\bk_3)}{(i \omega - \epsilon_{\bp})(i\omega - \epsilon_{\bp+\bq})}\,,
\end{align}
where the momentum transfer is $\bq = \bk_1 - \bk_3$ and the momenta relative to the Fermi momentum are $l_{\bp} = |\bp| - \kF$ and $l_{\bp + \bq} = |\bp + \bq| - \kF$.
After performing the integration over $\omega$, one finds
\begin{align}
  V_{\mathrm{ZS}}(\bk_1, \bk_3)
  \simeq 2V^2(\pi)\int_{|l_{\bp}|, |l_{\bp + \bq}|  <\Lambda} \frac{d^3\bp}{(2\pi)^3} \frac{\Theta (-\epsilon_{\bp})\Theta(\epsilon_{\bp + \bq})} {\epsilon_{\bp} - \epsilon_{\bp +\bq}}\,.
\end{align}
Here again, the integrand is singular at $\bq = - 2\bp$ due to the Kohn anomaly, so the integration is dominated by the contribution at $\bq \simeq -2\kF$.
This implies that the angle between $\bkhat_1$ and $\bkhat_3$ is close to $\pi$.
Therefore, we approximate the scattering amplitudes in the above expression with $V(\pi)$, which is the scattering amplitude of the back-to-back scattering at $\theta = \pi$ in the BCS channel.
In what follows, we focus on the problem of evaluating the following integral:
\begin{align}
    \calI &= \int \frac{d^3\bp}{(2\pi)^3} \frac{\Theta (-\epsilon_{\bp})\Theta(\epsilon_{\bp + \bq})} {\epsilon_{\bp} - \epsilon_{\bp +\bq}}\Theta(\Lambda - |l_{\bp}|) \Theta(\Lambda - |l_{\bp+\bq}|)\,.
    \label{eq:Idef}
\end{align}
The integral $\calI$ can be approximated as
\begin{align}
    \calI &\simeq \frac{\kF^2}{4\pi^2} \int_0^\infty dp \int_{-1}^{1} dz\, \frac{\Theta (-vl_{\bp})\Theta(vl_{\bp + \bq})} {vl_{\bp} - vl_{\bp +\bq}}\Theta(\Lambda - |l_{\bp}|) \Theta(\Lambda - |l_{\bp+\bq}|)\,,
\end{align}
where $z = \bphat \cdot \bqhat$.
Now, we perform the change of variables $p \to l_{\bp} $ and $z \to l_{\bp + \bq}$, and use the relations $d p = dl_{\bp}$ and $d z = \frac{\kF + l_{\bp + \bq}}{(\kF + l_{\bp})q} dl_{\bp + \bq} $.
We obtain
\begin{align}
    \calI = \frac{\nuF}{2q} \int_{-\kF}^\infty d l_{\bp} \int_{|q - p| - \kF}^{|q + p| - \kF} dl_{\bp+\bq}\, \frac{\kF + l_{\bp+\bq}}{\kF + l_{\bp}} \frac{\Theta (-l_{\bp})\Theta(l_{\bp + \bq})} {l_{\bp} - l_{\bp +\bq}}\Theta(\Lambda - |l_{\bp}|) \Theta(\Lambda - |l_{\bp+\bq}|)\,.
\end{align}
Below, we relabel $x = -l_{\bp}$ and $y = l_{\bp + \bq}$ and incorporate the step functions in the integration range.
We approximate the first term in the integrand as $(\kF + y) / (\kF - x) \simeq 1$ as the integration ranges of $x$ and $y$ are limited to small values compared to $\kF$.
Then, the integral $\calI$ can be evaluated as
\begin{align}
    \calI 
    &\simeq -\frac{\nuF}{2 q} \int_0^{\Lambda} d x \int_{\max(x- \tilde{q},\,0)}^{ \Lambda} dy\, \frac{1} {x+y}\,, \notag \\
    &= -\frac{\nuF}{2 q} \left \{ \theta(\Lambda - \tilde{q}) \left[2\Lambda \ln 2 + \frac12 \tilde{q} \ln\left(\frac{2\Lambda}{\tilde{q}} - 1\right) - \Lambda \ln \left(2 - \frac{\tilde{q}}{\Lambda}\right)\right]  + \theta(\tilde{q} - \Lambda) 2\Lambda \ln 2 \right\}\,,
    \label{eq:integI}
\end{align}
where we defined $\tilde{q} \equiv 2\kF - q$.
By taking the differentiation with respect to $\Lambda$, one gets
\begin{align}
    \Lambda \frac{d \calI}{d\Lambda}
    &= -\frac{\nuF \Lambda}{2 q}\left[2\ln2 - \theta(\Lambda - \tilde{q}) \ln\left(2 - \frac{\tilde{q}}{\Lambda} \right) \right]+ O\left(\frac{\Lambda^2}{\kF^2}\right) \,.
\end{align}
Since the integration is dominated by the contribution at $q = 2\kF$, we can safely assume that $\tilde{q} \ll \Lambda$ in the parameter range of interest.
Thus, the integral can be approximated as
\begin{align}
    \Lambda \frac{d \calI}{d\Lambda}
    &\simeq -\frac{\nuF \Lambda}{2 q}\ln2 + O\left(\frac{\Lambda^2}{\kF^2}\right) \,.
\end{align}
From Eq.~\eqref{eq:betaV}, one can evaluate the beta function for the ZS diagram $\beta_{\mathrm{ZS}} =- d V_{\mathrm{ZS}}/dt$ up to $O(\Lambda / \kF)$ as
\begin{align}
    \beta_{\mathrm{ZS}}
    =2V^2(\pi) \Lambda\frac{d \calI}{d\Lambda}
    = -\nuF V^2(\pi)  \ln 2\, \frac{\Lambda}{q}\,.
\end{align}
In the partial-wave expansion, the ZS beta function becomes
\begin{equation}
    \beta_{\mathrm{ZS},\ell}  = -\nuF  V^2(\pi) \frac{ \ln 2 }{2\ell+1} \frac{\Lambda}{\kF}\,,
\end{equation}
where we used the relation for $q(z) = \sqrt{2 (1-z)} \kF$:
\begin{equation}
    \frac12 \int_{-1}^1 dz\, P_{\ell}(z) \frac1{q(z)} = \frac{1}{2 \ell +1} \frac{1}{\kF}\,.
    \label{eq:qpartialwave}
\end{equation}
Note that this relation is exact.
We can verify that this integral above is dominated by the contribution from $z = -1$, which corresponds to $\theta = \pi$.
This justifies \textit{a posteriori} the approximation that we have limited the contributions of scattering amplitude to $\theta = \pi$ owing to the Kohn anomaly.

Likewise, one can calculate the ZS' diagram and obtains
\begin{align}
    \beta_{\mathrm{ZS'}}
    = \nuF V^2(\pi)  \ln 2\, \frac{\Lambda}{q'}\,.
\end{align}
In the partial-wave expansion, the ZS' beta function is
\begin{equation}
    \beta_{\mathrm{ZS'}, \ell} = (-1)^{\ell} \nuF  V^2(\pi) \frac{ \ln 2 }{2\ell+1} \frac{\Lambda}{\kF} \,,
\end{equation}
where we used the relation for $q'(z) = \sqrt{2 (1+z)} \kF$:
\begin{equation}
    \frac12 \int_{-1}^1 dz\, P_{\ell}(z) \frac1{q'(z)} = \frac{(-1)^{\ell}}{2 \ell +1} \frac{1}{\kF}\,.
\end{equation}

Before solving the RG equation with the ZS beta function, let us clarify how our result differs from the previous evaluation of $\beta_{\mathrm{ZS}}$.
In Appendix B of Ref.~\cite{Shankar:1993pf}, the asymptotic values of the ZS beta function was evaluated in the limits $\Lambda \to 0$ and $\ell \to \infty$, while keeping either $1/(\Lambda \ell^2) \to \infty$ or $\Lambda \ell^2 \to \infty$.
The limit considered here corresponds to the latter case, $\Lambda \ell^2 \to \infty$, since we keep $\Lambda / \ell$ constant.
In Ref.~\cite{Shankar:1993pf}, the partial-wave expansion of $\calI$~\eqref{eq:Idef} in the limit $\Lambda \ell^2 \to \infty$ yields $\calI_{\ell} \sim \Lambda^0/\ell^4$.
Meanwhile, our result is $\calI_{\ell} \sim \Lambda^1 / \ell$.
At first glance these appear to conflict, but they are, in fact, compatible.
Indeed, the same term $\calI_{\ell} \sim \Lambda^0 / \ell^4$ is already contained in Eq.~\eqref{eq:integI}, originating from the integrand $\tilde{q} \ln \tilde{q} = (2\kF -q) \ln (2\kF - q)$.
This $q$-dependence underlies the conventional KL irreducible interaction, which scales like $1/\ell^4$.
When one differentiates the $\Lambda^0 / \ell^4$ term with respect to $\Lambda$, it vanishes, thus making no contribution to the beta function.
Consequently, the subleading term $\calI_{\ell} \sim \Lambda^1 / \ell$ becomes the leading term for the beta function.
Hence, our calculation and that of Ref.~\cite{Shankar:1993pf} are indeed consistent.
Whereas reference~\cite{Shankar:1993pf} focused on the leading contribution in $(\Lambda / \kF)^0$, which vanishes upon the differentiation with respect to $\Lambda$, we computed the subleading contribution of the order of $(\Lambda/ \kF)^1$.
Additionally, reference~\cite{Shankar:1993pf} also derived an expression in the other limit $1/(\Lambda \ell^2) \to \infty$, which contributes at $O(\Lambda^{7/4}/\kF^{7/4})$.
This is subleading compared to our $O(\Lambda/\kF)$ term and can therefore be neglected.

\section{Enhanced exponent in the Kohn-Luttinger mechanism}
\label{sec:enhance}

In this section, we solve the RG equation including the ZS beta function, and one sees the enhanced exponent in the Kohn-Luttinger effect from the solution.
The RG equation including the ZS beta function is 
\begin{align}
    \frac{d V_{\ell}(t)}{dt}
    = - \nuF V^2(\pi;t) \frac{\ln 2}{2\ell + 1} e^{-t} - \frac{\nuF}{2} V_{\ell}^2(t)\,.
\end{align}
Here the scale parameter is $t = -\ln (\Lambda / \mu) > 0$ and $\mu = \kF^2 / (2m)$.
In principle, $V(\pi; t)$ also has $t$ dependence and thus runs with $t$, but we neglect its $t$-dependence here since the whole contribution from $\beta_{\mathrm{ZS}}$ is centered around $t \simeq 0$, and approximate it as $V(\pi; t) \simeq V(\pi; t=0)$.
By absorbing the factor $\nuF$ in the definition of $V_{\ell}$ and $V(\pi)$ to make them dimensionless, i.e.\ $V \to \hat{V} = \nuF V$, one can rewrite the RG equation as
\begin{equation}
    \frac{d \hat{V}_{\ell}(t) }{dt} = - \hat{V}^2(\pi) \frac{\ln 2}{2\ell+1} e^{-t}  - \frac{\hat{V}_{\ell}^2(t)}{2}\,.
\end{equation}
In the rest of the paper, we use the dimensionless $V$ and omit the hat in $\hat{V}$.
The solution of this RG equation is
\begin{align}
    \label{eq:sol}
    V_{\ell}(t) &= c_{\ell} e^{-t/2} \frac{\mathcal{X}}{\mathcal{Y}}\,, \\
    \label{eq:solX}
    \mathcal{X} &= \left[V_{\ell}(0) J_0\left(c_{\ell} \right) - c_{\ell}J_1\left(c_{\ell} \right) \right] Y_1\left(e^{-t/2} c_{\ell} \right)
    -\left[V_{\ell}(0) Y_0\left( c_{\ell} \right) - c_{\ell} Y_1\left(c_{\ell} \right) \right] J_1\left(e^{-t/2}c_{\ell}\right)\,, \\
    \label{eq:solY}
    \mathcal{Y} &= \left[V_{\ell}(0) J_0\left( c_{\ell} \right) - c_{\ell} J_1\left( c_{\ell} \right) \right] Y_0\left(e^{-t/2} c_{\ell} \right)
    -\left[V_{\ell}(0) Y_0\left(c_{\ell} \right) - c_{\ell} Y_1\left(c_{\ell} \right) \right] J_0\left(e^{-t/2} c_{\ell}\right)\,, 
\end{align}
where we defined $c_{\ell} \equiv \frac{V(\pi)\sqrt{2\ln2}}{ \sqrt{2\ell+1}}$.
Since $c_{\ell} \ll 1$ at large $\ell$, we expand $\mathcal{Y}$ to locate the BCS singularity $\mathcal{Y} = 0$:
\begin{align}
    \mathcal{Y} &\simeq
    \left(V_{\ell}(0) - \frac{c_{\ell}^2}{2} \right) \frac{2}{\pi} \left( \gamma + \ln \frac{c_{\ell}}{2} - \frac{t}{2} \right)
    -\left[V_{\ell}(0) \frac{2}{\pi}\left(\gamma + \ln\frac{c_{\ell}}{2} \right) + \frac{2}{\pi} \right]\,, \notag \\
    &=-\frac2{\pi}\left[1 +O(c_\ell^2) \right] -\left(V_{\ell}(0) -V^2(\pi)  \frac{\ln2 }{2\ell+1} \right) \frac{t}{\pi}\,.
    \label{eq:calY}
\end{align}
In order for $V_{\ell}(t)$ to be singular, the second term has to be negative, namely, $V_{\ell}(0) - V^2(\pi) \frac{\ln2 }{2\ell+1} < 0$.
This is fulfilled when $V_{\ell}(0) \leq 0$.
When the initial coupling at the beginning of the RG flow vanishes, $V_{\ell}(t)$ always reaches the singularity regardless of the sign of $V(\pi)$.
This is the RG version of the KL mechanism.
In obtaining the above expansion, we used the following formulas for the Bessel functions for $x\ll 1$,
\begin{align}
    J_0(x) &\simeq 1 - \frac{x^2}{4} + O(x^4)\,,\\
    J_1(x) & \simeq \frac{x}{2} - \frac{x^3}{8} + O(x^5)\,,\\
    Y_0(x) & \simeq \frac{2}{\pi} \left(\gamma + \ln\frac{x}{2} \right) J_0(x) + \frac{x^2}{2\pi} + O(x^4) \,,\\
    Y_1(x) & \simeq - \frac{2}{\pi x} + \frac{2}{\pi} \ln \frac{x}{2} J_1(x) - \frac{1}{\pi}(1 - 2\gamma) \frac{x}{2}\,.
\end{align}
By keeping only the leading-order terms in the numerator and denominator, respectively, one can locate the BCS singularity at
\begin{align}
    t
    &\simeq -\frac{2}{V_{\ell}(0) - V^2(\pi) \frac{\ln2 }{2 \ell +1}  }\,.
\end{align}
Note that the arguments of $V$ in the first and second terms of the denominator refer to $t=0$ and $\theta = \pi$, respectively.

Let us find the pairing gap from the KL mechanism from the above expression by specifying the initial condition for $V_{\ell}(t)$ at $t=0$.
As in the argument of the ordinary KL mechanism for the repulsive interaction, reviewed in the previous section, one obtains at a one loop order in the perturbation theory at large $\ell$
\begin{align}
    V_{\ell}(0) \sim - \frac{c}{\ell^4}\,,
\end{align}
where the coefficient $c$ is a positive number of the order of 1.
Therefore, for very large $\ell \gg 1$, one obtains the pairing gap
\begin{align}
    \ln \left(\frac{\Delta}{\mu}\right) 
    &\simeq \frac{2}{- \frac{c}{\ell^4} - V^2(\pi) \frac{\ln 2}{2\ell+1}}\,,\notag \\
    & \simeq -\frac{2(2\ell + 1)}{V^2(\pi) \ln 2} \propto -\ell\,.
\end{align}
For very large $\ell$, based on the RG argument, we find that the gap $\Delta$ arising from the KL mechanism has the behavior $\ln \Delta \propto -\ell$ instead of $\ln \Delta \propto - \ell^4$ as has been known for a long time.

Note that this result is specific to the three-dimensional (3D) system, and the scaling may depend on the dimensionality.
In the 2D case, the integration measure for the angular integral is different from 3D, and the eigenfunctions for the angular momentum become $\cos(\ell\theta)$ in 2D.
Therefore, the derivation of the beta function from the ZS diagram in the previous section gets altered, and the $\ell$ dependence may well change.

\section{Spin triplet p-wave superfluid gap}
\label{sec:tripletpwave}

In this section, we exemplify the KL effect with the $\ell=1$ case.
We consider the spin-triplet p-wave superfluid gap in the presence of the s-wave bare repulsion.
As we will see, the LO perturbation theory fails at $\ell \geq 1$.
This implies that the conventional argument of the KL effect based on the LO perturbation theory may not be enough to capture the pairing in the higher partial wave channel.
We claim that the RG approach can effectively take this into account in a concise manner.

\subsection{Pairing gap from the renormalization group}

So far, we have neglected the spin degrees of freedom.
Here, let us extend the RG equation to include the spin.
Generally, one can decompose $V$ into the spin singlet and triplet components, $\Vsinglet$ and $\Vtriplet$, respectively:
\begin{equation}
    (V_{\ell})_{\sigma_1 \sigma_2 \sigma_3 \sigma_4}
    = \Vsinglet_{\ell} (\delta_{\sigma_1 \sigma_3} \delta_{\sigma_2 \sigma_4} - \delta_{\sigma_1 \sigma_4} \delta_{\sigma_2 \sigma_3})
    + \Vtriplet_{\ell} (\delta_{\sigma_1 \sigma_3} \delta_{\sigma_2 \sigma_4} + \delta_{\sigma_1 \sigma_4} \delta_{\sigma_2 \sigma_3})\,.
    \label{eq:stdecomp}
\end{equation}
The RG equations for $\Vsinglet_{\ell}$ and $\Vtriplet_{\ell}$ with taking into account the ZS and ZS' beta functions as well as spin degrees of freedom are as follows.
When $\ell$ is odd,
\begin{align}
    \frac{d \Vsinglet_{\ell}}{dt} &= -  \left(\Vsinglet_{\ell}\right)^2 \,,\\
    \frac{d \Vtriplet_{\ell}}{dt} &= -  \left(\Vtriplet_{\ell}\right)^2 -  \left[\left(\Vsinglet(\pi) \right)^2 
 + 5 \left(\Vtriplet(\pi) \right)^2 + 2 \Vsinglet(\pi) \Vtriplet(\pi) \right] \frac{ \ln 2 }{2\ell+1} e^{-t}\,.
 \label{eq:RGtriplet}
\end{align}
When $\ell$ is even,
\begin{align}
    \frac{d \Vsinglet_{\ell}}{dt} &= - \left(\Vsinglet_{\ell}\right)^2 +  \left[\left(\Vsinglet(\pi) \right)^2 
 - 3 \left(\Vtriplet(\pi) \right)^2 - 6 \Vsinglet(\pi) \Vtriplet(\pi) \right] \frac{ \ln 2 }{2\ell+1} e^{-t}\,,\\
    \frac{d \Vtriplet_{\ell}}{dt} &= - \left(\Vtriplet_{\ell}\right)^2 \,.
\end{align}
We provide the detailed derivation in Appendix~\ref{sec:appendix}.
The solution to these equations is as well given by Eqs.~(\ref{eq:sol}--\ref{eq:solY}).

Let us now consider the formation of the BCS instability in the spin triplet p-wave channel, which is described by Eq.~\eqref{eq:RGtriplet} above.
We consider a field theory with the repulsive contact interaction in s-wave:
\begin{equation}
    u_{\sigma_1 \sigma_2 \sigma_3 \sigma_4}(l_1, l_2, l_3, l_4) = \frac{4\pi a}{m} (\delta_{\sigma_1 \sigma_3} \delta_{\sigma_2 \sigma_4} - \delta_{\sigma_1 \sigma_4} \delta_{\sigma_2 \sigma_3})\,,
\end{equation}
where $a>0$ is the scattering length.
The pairing gap read out from the solution of the RG equation is 
\begin{align}
    \ln \left(\frac{\Delta}{\mu}\right) = \frac{1}{\Vtriplet_{\ell = 1}(0) - \left[\left(\Vsinglet(\pi) \right)^2 + 5 \left(\Vtriplet(\pi) \right)^2 + 2 \Vsinglet(\pi) \Vtriplet(\pi) \right] \frac{\ln2 }{3}  }\,.
    \label{eq:tripletgap}
\end{align}

Now, let us determine the coupling functions that goes into the gap expression up to the leading order in the perturbation theory.
The initial conditions for the RG flow is specified at $t = 0$, or $\Lambda = \kF$.
By substituting $\Lambda = \kF$ in Eq.~\eqref{eq:Idef}, one obtains
\begin{equation}
    \Vtriplet(\theta; t=0) = \nuF \left(\frac{4\pi a}{m}\right)^2 \int \frac{d^3\bp}{(2\pi)^3} \frac{2\Theta (-\epsilon_{\bp})\Theta(\epsilon_{\bp + \bq})} {\epsilon_{\bp} - \epsilon_{\bp +\bq}}\Theta(\kF - |l_{\bp}|) \Theta(\kF - |l_{\bp+\bq}|)\,.
\end{equation}
Now, the integrand is the same as the one in the Lindhard function~\eqref{eq:lindhard} by resorting to the relation $\Theta(|\bp + \bq| - \kF) - \Theta(|\bp| - \kF) = 2\Theta(|\bp + \bq| - \kF) \Theta(\kF - |\bp|)$, which holds for the integrand that is antisymmetric with respect to the exchange $|\bp| \leftrightarrow |\bp + \bq|$.
Therefore,
\begin{equation}
    \Vtriplet(\theta; t=0) = \nuF \frac{4\pi a}{m}\Pi(\omega=0, \sqrt{2(1-\cos\theta)}\kF)\,,
\end{equation}
and in the partial wave expansion, it becomes
\begin{align}
    \Vtriplet_{\ell = 1}(t=0) &= \frac12 \int_{-1}^{1} d \cos\theta\, P_{\ell = 1}(\cos\theta)\Vtriplet(\theta; t=0)
    = -\frac{2 \ln 2 - 1}5 \lambda^2\,,
    \label{eq:Vtl1}
\end{align}
where the perturbative expansion parameter commonly used in this context is defined as $\lambda \equiv 2 \kF a / \pi$.
For the coupling function in the back-to-back scattering at $\theta = \pi$, one finds at the lowest order in the perturbation theory:
\begin{align}
    \label{eq:Vspi}
    \Vsinglet(\theta = \pi) &= \nuF \frac{4\pi a}{m}= \lambda\,,\\
    \label{eq:Vtpi}
    \Vtriplet(\theta = \pi) &= 0\,.
\end{align}

The pairing gap is obtained by substituting Eqs.~(\ref{eq:Vtl1}, \ref{eq:Vspi}, \ref{eq:Vtpi}) into Eq.~\eqref{eq:tripletgap}:
\begin{align}
    \Delta = \mu \exp \left(\frac{1}{-\frac{2 \ln 2 - 1}{5}\lambda^2 - \frac{\ln 2}3 \lambda^2}\right)\,.
    \label{eq:RGgap}
\end{align}
The contribution of the ZS and ZS' beta functions appears as in the second term in the exponent's denominator.
When this term is neglected, one recovers the one-loop perturbative expression from the Bethe-Salpeter equation, as one can see in Eq.~\eqref{eq:gammapert}.
This implies that modification takes place in the Bethe-Salpeter equation instead in the KL induced interaction.
The RG equation with the ZS and ZS' beta functions modifies the Bethe-Salpeter equation shown in the top panel of Fig.~\ref{fig:KL}; the RG equation is capable of summing the BCS, ZS, and ZS' diagrams in a nested way, while the Bethe-Salpeter equation can only sum up the BCS diagrams.

\subsection{Comparison with the perturbation theory}

We define $\Gamma$ as $\Delta \sim \mu \exp(1/\nuF \Gamma)$, and below we compare the behavior of $\Gamma$ in the RG approach $\Gamma_{\mathrm{RG}}$ and the perturbation theory $\Gamma_{\mathrm{pert}}$.
On the one hand, in the RG approach, one finds from Eq.~\eqref{eq:RGgap}
\begin{align}
    \nuF \Gamma_{\mathrm{RG}}
    &= -\frac{2 \ln 2 - 1}{5}\lambda^2 - \frac{\ln 2}3 \lambda^2 + O(\lambda^3)\,,\notag\\
    &\simeq - 0.31 \lambda^2 \,.
    \label{eq:gammaRG}
\end{align}
On the other hand, in the perturbation theory, the leading-order coefficient was calculated by Fay and Layzer a long ago~\cite{PhysRevLett.20.187} and also by Kagan and Chubukov~\cite{1988JETPL..47..614K}.
The higher-order corrections to the leading-order coefficient were obtained up to $O(\lambda^4)$ in Ref.~\cite{2000JETP...90..861E} numerically.
The results in perturbation theory up to $O(\lambda^4)$ reads
\begin{align}
    \nuF \Gamma_{\mathrm{pert}} &= - \frac{2\ln 2 - 1}{5} \lambda^2 - 0.33 \lambda^3 - 0.26 \lambda^4 + O(\lambda^5)\,,\notag\\
    &\simeq - 0.077 \lambda^2 - 0.33 \lambda^3 - 0.26 \lambda^4\,.
    \label{eq:gammapert}
\end{align}
The leading-order coefficient is anomalously small compared to the subleading coefficients, and the perturbative series converges poorly up to the next to leading order (NLO).
When one goes to one order higher up to next to NLO (NNLO), then the convergence improves.
The conventional description given in the literature is that the higher-order terms have a strong angular dependence compared to the leading-order terms, so the higher-order coefficients of $\Gamma$ are enhanced in the partial wave expansion.
The diagram containing the BCS diagram as a subdiagram gets an enhancement in the numerical coefficients.
In the next subsection, we will clarify that such BCS loops always appear with the Kohn anomaly at two-loop order.

We note in passing that attractive s-wave pairing does not exhibit the same convergence issue.
In particular, the p-wave pairing series starts at $O(\lambda^2)$, whereas the s-wave pairing series begins at $O(\lambda)$.
Indeed, for s-wave pairing, the tree-level amplitude at $O(\lambda)$ is nonzero, and the Gor'kov--Melik-Barkhudarov correction to the BCS result---which account for induced interactions---enters at $O(\lambda^2)$~\cite{Gorkov:1961}.
Recently, the s-wave pairing gap was numerically computed up to $O(\lambda^3)$ in Ref.~\cite{Beane:2024tmd}, with no apparent convergence issues observed through that order.

\begin{figure}
    \centering
    \includegraphics[width=0.5\linewidth]{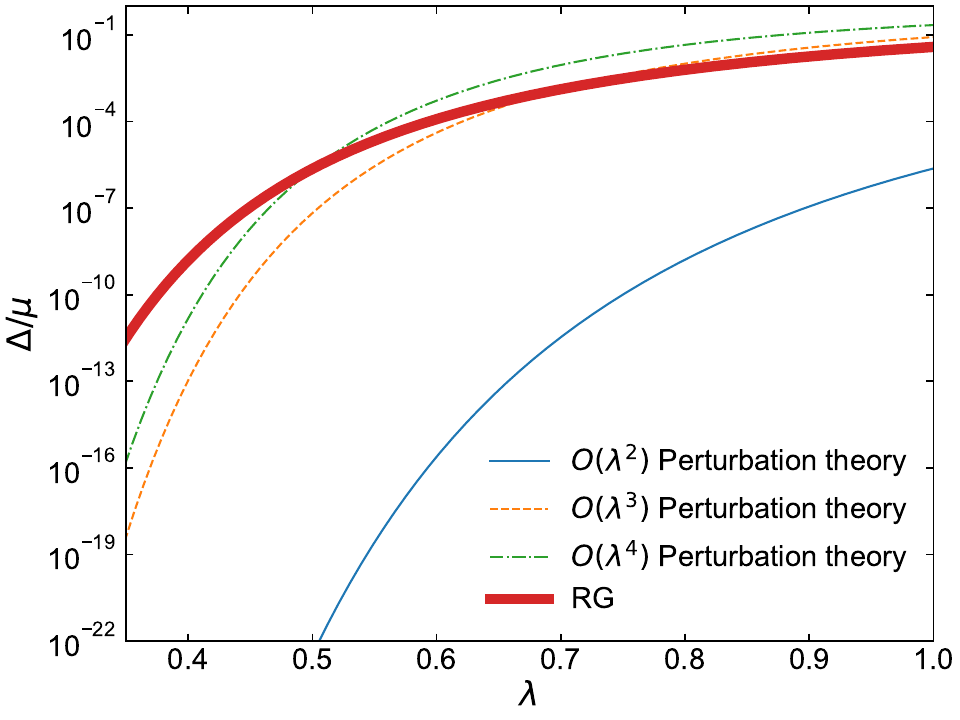}
    \caption{Comparison of the gap evaluated in the perturbation theory and in the RG approach.}
    \label{fig:gap}
\end{figure}
In Fig.~\ref{fig:gap}, we compare the behavior of the gap parameter $\Delta / \mu$ as a function of expansion parameter $\lambda$ evaluated in the perturbation theory and the RG approach.
As one can clearly see, the LO perturbation theory at $O(\lambda^2)$ completely fails as they are orders of magnitude smaller compared to the NLO ($O(\lambda^3)$) and NNLO ($O(\lambda^4)$) results.
The RG approach matches well with the NLO result from the perturbation theory at a moderately large value of $\lambda \gtrsim 0.5$, where the gap parameter is relatively large, and thus measurable size in reality.
This indicates that already at the level of the LO, the RG approach can virtually capture the NLO contribution in the perturbation theory.

Our intension here is to point out the issue related to the convergence of the perturbative series, which should formally be valid as long as $\lambda$ is less than one.
The convergence issue becomes more noticeable as $\lambda$ gets large, and thus, the plot range of the figure is limited to $\lambda > 0.35$.
This choice is also motivated by that the solution of the Bethe-Salpeter equation should work well in the small-$\lambda$ region, and the pairing gap becomes exponentially suppressed and too small to be of phenomenological interest below $\lambda < 0.4$.

\subsection{Possible explanation}
\begin{figure}
    \centering
    \includegraphics[width=0.6\columnwidth]{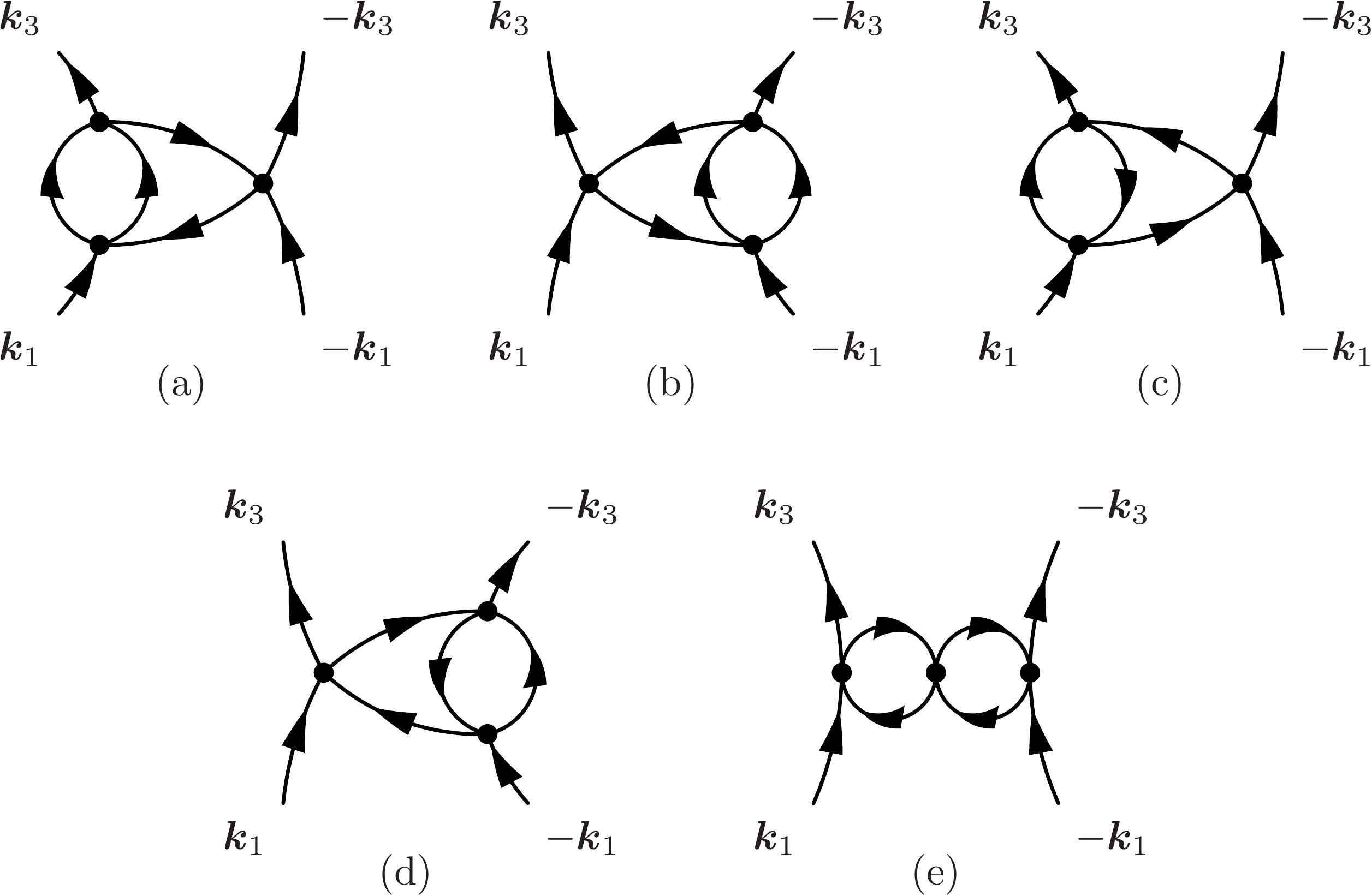}
    \caption{The skeleton diagrams that contribute to the two-loop order in the perturbation theory. There are also diagrams with $\bk_3 \leftrightarrow -\bk_3$.}
    \label{fig:3rd}
\end{figure}

We consider the diagram at the two-loop order in the perturbation theory that contains the BCS loops as a subdiagram.
Such a diagram is evaluated numerically to have a larger value compared to those that do not contain the BCS loops as a subdiagram by orders of magnitude~\cite{2000JETP...90..861E, Beane:2024tmd}.
We list the skeleton diagrams that contribute to the third order in the perturbation theory in Fig.~\ref{fig:3rd}.
Figs.~\ref{fig:3rd}~(a) and (b) contain the BCS loops as a subdiagram.
Let us evaluate $\Pi_{\mathrm{ppph}}$ that arises from the diagram (a):
\begin{align}
    \Pi_{\mathrm{ppph}}(E, \bk_1, \bk_3) = \int \frac{d^4 p}{(2\pi)^4} \int \frac{d^4 p'}{(2\pi)^4} G_0 (E + p_0, \bp) G_0(p_0' - p_0, \bp' - \bp) G_0(p_0', \bp' - \bk_1) G_0(p_0', \bp' - \bk_3)\,,
\end{align}
where the propagator is
\begin{equation}
    G_0(k_0, \bk) = \frac{i}{k_0 - \omega_k + i0^+} - 2\pi \delta(k_0 - \omega_k) \theta(\kF - k)\,.
\end{equation}
After performing the energy integral, one obtains~\cite{Beane:2024tmd}
\begin{align}
    \Pi_{\mathrm{ppph}}(E, \bk_1, \bk_3) &= \int \frac{d^3 \bp}{(2\pi)^3} \int \frac{d^3 \bp'}{(2\pi)^3} \bigg[\frac{\theta(\kF - q_3) - \theta(p - \kF)}{\omega_{q_1} - \omega_{q_2}} \left( \frac{\theta(\kF - q_2)}{\omega_{q_2} + E - \omega_p - \omega_{q_3}} - \frac{\theta(\kF - q_1)}{\omega_{q_1} + E - \omega_p - \omega_{q_3}} \right) \notag \\
    &\qquad\qquad\qquad\qquad\qquad - \theta(\kF - p) \theta(\kF - q_3) \frac{1}{(-\omega_{q_1} - E + \omega_p + \omega_{q_3})(-\omega_{q_2} - E + \omega_p + \omega_{q_3})} \bigg]\,,
\end{align}
where the vectors are defined as $\bq_1 = \bp' - \bk_3$, $\bq_2 = \bp' - \bk_1$, and $\bq_3 = \bp' - \bp$.

Below, we focus on the IR structure of the first integral in the square bracket.
By using the symmetry $\omega_1 \leftrightarrow \omega_2$, the integral reduces to
\begin{equation}
    \mathcal{J} = \int \frac{d^3 \bp}{(2\pi)^3} \int \frac{d^3 \bp'}{(2\pi)^3} \frac{\theta(\kF - q_3) - \theta(p - \kF)}{\omega_{q_1} - \omega_{q_2}} \frac{\theta(\kF - q_2)}{\omega_{q_2} + E - \omega_p - \omega_{q_3}}\,.
\end{equation}
When the conditions $\omega_{q_1} - \omega_{q_2} = 0$, $|\bk_1 - \bk_3| = 2\kF$, and $\omega_{q_2} +E - \omega_p - \omega_{q_3} = 0$ are fulfilled simultaneously, this will lead to the divergent integrand.
The first and second conditions correspond to the Kohn anomaly, and the third condition corresponds to the BCS instability.
Such case is achieved within kinematics: $p \sim \kF$, $p' \sim 0$, $\bphat \cdot \bphat' \sim -1$, and $\bphat \cdot \bkhat_1 \sim 0$.
Although the integrand is divergent, the integral itself does not have non-integrable singularities since the divergence is canceled by the phase space factor in the integral measure, so this is amenable to the numerical integral.
The integral that contains the singular part in the integrand is
\begin{align}
    \mathcal{J} &\sim 
    \int^{\kF} p^2 dp \int \frac{d\bphat}{4\pi}
    \int_0 p'^2 dp' \int \frac{d\bphat'}{4\pi} \,\frac{1}{2 m E + \kF^2 - 2p^2 + 2\bp' \cdot (\bk_1 - \bk_3)}
    \frac{1}{2 \bp' \cdot (\bk_1 - \bk_3)} \,, \notag \\
    &\sim
    \int^{\kF} p^2 dp\, \frac{1}{2 m E + \kF^2 - 2p^2}
    \int_0 p'^2 dp'\, \frac{1}{2 p' |\bk_1 - \bk_3|} \,.
\end{align}
From the first to the second line, we neglected the term $2\bp' \cdot (\bk_1 - \bk_3)$ in the denominator of the first fraction since $p' \sim 0$ in this kinematics.
As one can see, the integral $\mathcal{J}$ can be factored into two pieces; the former integral $\sim \ln(2 m E - \kF^2)$ corresponds to the BCS instability, and the latter integral corresponds to the Kohn anomaly.
The latter integral also scales as $1/\ell$ in the partial wave expansion because the partial wave expansion of $1/|\bk_1 - \bk_3|$ is exactly the same as Eq.~\eqref{eq:qpartialwave}.
Therefore, we also see $1/\ell$ behavior in the perturbation theory.
When $\lambda$ takes a fixed moderate value and $\ell \gg 1$, the contribution from the one-loop diagrams, $\sim \lambda^2 / \ell^4$, is smaller than the contribution from the two-loop diagrams, $\sim \lambda^3 / \ell$.
Such enhancement at the two-loop order was also observed in the KL mechanism in 2D~\cite{PhysRevB.48.1097}.
The enhancement in the 2D system may also be captured within the RG approach.
We leave this for future work.

\section{Conclusions and outlook}

In this work, we revisited the renormalization group analysis of the Kohn-Luttinger mechanism.
Our primary focus was on extracting the singularity in the scattering amplitude under BCS kinematics, where the BCS instability occurs.
Traditionally, renormalization-group analyses retain only relevant or marginal operators when studying the flow toward a fixed point.
However, to address the BCS instability, the contributions from irrelevant operators must also be included in the beta function.
By solving the renormalization group equation accordingly, we obtained the gap parameter.
The key findings are as follows:
\begin{itemize}
    \item Enhanced exponent: The exponent of the gap from the Kohn-Luttinger mechanism scales as $\propto -\ell$ instead of $\propto -\ell^4$, thus increasing the gap compared to the conventional $\ell^4$ scaling.
    \item Limitations of leading-order perturbation theory: Solving the Bethe-Salpeter equation at leading order may not fully capture the Kohn-Luttinger mechanism.
    The poor convergence observed in perturbation series indicates the need for some form of resummation.
    \item Renormalization-group approach and subleading effects: The renormalization-group method can incorporate subleading corrections in perturbation theory by using only the leading-order terms as inputs to the RG flow.
\end{itemize}

A further test of this enhanced exponent in the Kohn-Luttinger mechanism can be conducted within perturbation theory at next-to-leading order by examining the $\ell$ dependence of the integral
\begin{equation}
    \int_{-1}^{1} d\cos\theta \, P_{\ell}(\cos\theta) \left[\Pi_{\mathrm{ppph}}^{\mathrm{(a)}}(\bk_1, \bk_3) + \Pi_{\mathrm{phph}}^{\mathrm{(c)}}(\bk_1, \bk_3)\right]\,,
\end{equation}
valid for odd $\ell$.
We retain only $\Pi_{\mathrm{ppph}}^{\mathrm{(a)}}$ and $\Pi_{\mathrm{phph}}^{\mathrm{(c)}}$ because they can be mapped to $\Pi_{\mathrm{ppph}}^{\mathrm{(b)}}$ and $\Pi_{\mathrm{phph}}^{\mathrm{(d)}}$ via $\bk_1 \to -\bk_1$ and $\bk_3 \to - \bk_3$.
Additionally, $\Pi_{\mathrm{phph}}^{\mathrm{(e)}} = - \Pi^2$ does not contribute in the odd partial-wave sector.
At present, there are no known analytical expressions for $\Pi_{\mathrm{ppph}}^{\mathrm{(a)}}$ and $\Pi_{\mathrm{phph}}^{\mathrm{(c)}}$, so they can only be evaluated numerically.
If the above renormalization-group reasoning holds, this integral should scale as $1/\ell$, not $1/\ell^4$.

A key advantage of the renormalization-group approach is that it can effectively reproduce the two-loop solution of the Bethe-Salpeter equation by employing only the one-loop perturbative inputs, provided the expansion parameter is not too small.
Given the portability of the renormalization-group method, one promising application of this amplified Kohn-Luttinger mechanism is to nuclear and quark matter, which may exist in neutron stars.
A recent one-loop perturbative analysis can be found in Ref.~\cite{Kumamoto:2024muw}; going beyond leading order via renormalization-group method could offer deeper insights.
We note that our results extend beyond the simple BCS approximation and may thus alter the conventional picture, e.g., in Ref.~\cite{Schwenk:2003bc}.
Further, enhanced pairing in quark matter may impact the phase structure inside neutron stars~\cite{Fujimoto:2019sxg}.
Future work will apply this methodology to nuclear matter and quark matter along the lines suggested in Ref.~\cite{Kumamoto:2024muw}.

\begin{acknowledgments}
The author gratefully acknowledges
Andrey Chubukov,
Roland Farrell,
and
Sanjay Reddy for valuable correspondence and discussions.
I would like to thank
Silas Beane,
David Kaplan,
Mia~Kumamoto,
William~Marshall,
and 
Achim Schwenk
for helpful conversations at various stages of this work.

This work is in part supported by the Institute for Nuclear Theory's U.S. DOE Grant No.\ DE-FG02-00ER41132, Japan Science and Technology Agency (JST) as part of Adopting Sustainable Partnerships for Innovative Research Ecosystem (ASPIRE), Grant No.\ JPMJAP2318, the National Science Foundation Grant No. PHY-2020275, and the Heising-Simons Foundation Grant 2017-228.

\end{acknowledgments}

\appendix

\section{Including the spin degrees of freedom in the renormalization group equation}
\label{sec:appendix}

In this appendix, we derive the RG equation including the spin degrees of freedom.

As in Eq.~\eqref{eq:stdecomp}, one can decompose the coupling function into the spin singlet and triplet components, $\usinglet$ and $\utriplet$, respectively
\begin{equation}
    u_{1234}
    = \usinglet (\delta_{13} \delta_{24} - \delta_{14} \delta_{23})
    + \utriplet (\delta_{13} \delta_{24} + \delta_{14} \delta_{23})\,,
\end{equation}
where the subscript $1234$ refers to the spins $\sigma_1 \sigma_2 \sigma_3 \sigma_4$ of the particles $1$, $2$, $3$, and $4$.
The one-loop corrections including the spins are
\begin{align}
    (\delta V_{\mathrm{ZS}})_{1234}(\bk_1, \bk_3)
    &=  \sum_{\alpha,\beta=\uparrow,\downarrow}\int \frac{d^4 P}{(2\pi)^4}
    \frac{u_{1 \alpha 3 \beta}(\bk_1, \bp, \bk_3, \bp + \bq) u_{\beta 2 \alpha 4}(-\bp, -\bk_1, -(\bp+\bq), -\bk_3)}{(i \omega - \epsilon_{\bp})(i\omega - \epsilon_{\bp+\bq})}\,,\\
    (\delta V_{\mathrm{ZS'}})_{1234}(\bk_1, \bk_3)
    &=  -\sum_{\alpha,\beta=\uparrow,\downarrow}\int \frac{d^4 P}{(2\pi)^4}
    \frac{u_{1 \alpha 4 \beta}(\bk_1, \bp, -\bk_3, \bp + \bq') u_{\beta 2 \alpha 3}(-\bp, -\bk_1, -(\bp+\bq'), \bk_3)}{(i \omega - \epsilon_{\bp})(i\omega - \epsilon_{\bp+\bq'})}\,,\\
    (\delta V_{\mathrm{BCS}})_{1234}(\bk_1, \bk_3)
    &= - \sum_{\alpha,\beta=\uparrow,\downarrow} \frac12 \int \frac{d^4 P}{(2\pi)^4}
    \frac{u_{1 2 \alpha \beta}(\bk_1, -\bk_1, \bp, -\bp) u_{\alpha \beta 3 4}(\bp, -\bp, \bk_3, -\bk_3)}{(i \omega - \epsilon_{\bp})(-i\omega - \epsilon_{-\bp})}\,.
\end{align}
The coupling functions can be rewritten in terms of the singlet and triplet components as
\begin{align}
    \sum_{\alpha, \beta = \uparrow, \downarrow} u_{1 \alpha 3 \beta} u_{\beta 2 \alpha 4} &= 4 \utriplet \left( \usinglet + \utriplet \right) \delta_{13}\delta_{24} + \left( \usinglet - \utriplet \right)^2 \delta_{14} \delta_{23}\,,\\
    \sum_{\alpha, \beta = \uparrow, \downarrow} u_{1 \alpha 4 \beta} u_{\beta 2 \alpha 3} &= \left( \usinglet - \utriplet \right)^2 \delta_{13} \delta_{24} + 4 \utriplet \left( \usinglet + \utriplet \right) \delta_{14}\delta_{23}\,,\\
    \sum_{\alpha, \beta = \uparrow, \downarrow} u_{1 2 \alpha \beta} u_{\alpha \beta 3 4} &= 2 \left( \usinglet \right)^2 (\delta_{13} \delta_{24} - \delta_{14} \delta_{23})
    +  2 \left( \utriplet \right)^2 (\delta_{13} \delta_{24} + \delta_{14} \delta_{23})\,.
\end{align}
The beta functions are
\begin{align}
    (\beta_{\mathrm{ZS}, \ell})_{1234} &= -4 \nuF \Vtriplet(\pi) \left[ \Vsinglet(\pi) + \Vtriplet(\pi) \right] \frac{\ln2}{2\ell+1} \frac{\Lambda}{\kF} \delta_{13}\delta_{24}\notag \\
    &\qquad - \nuF \left[ \Vsinglet(\pi) - \Vtriplet(\pi) \right]^2 \frac{\ln2}{2\ell+1} \frac{\Lambda}{\kF} \delta_{14} \delta_{23}\,,\\
    (\beta_{\mathrm{ZS'}, \ell})_{1234}
    &= (-1)^{\ell} \nuF \left[ \Vsinglet(\pi) - \Vtriplet(\pi) \right]^2 \frac{\ln2}{2\ell+1} \frac{\Lambda}{\kF} \delta_{13} \delta_{24}\notag\\  
    &\qquad + 4 (-1)^{\ell}  \nuF \Vtriplet(\pi) \left[ \Vsinglet(\pi) + \Vtriplet(\pi) \right] \frac{\ln2}{2\ell+1} \frac{\Lambda}{\kF} \delta_{14}\delta_{23}\,, \\
    (\beta_{\mathrm{BCS}, \ell})_{1234} &= - \nuF \left( \Vsinglet_{\ell} \right)^2 (\delta_{13} \delta_{24} - \delta_{14} \delta_{23})
    - \nuF \left( \Vtriplet_{\ell} \right)^2 (\delta_{13} \delta_{24} + \delta_{14} \delta_{23})\,,
\end{align}
Then, the RG equation becomes
\begin{align}
    &\frac{d\Vsinglet_{\ell}}{dt} (\delta_{13} \delta_{24} - \delta_{14} \delta_{23})
    + \frac{d\Vtriplet_{\ell}}{dt} (\delta_{13} \delta_{24} + \delta_{14} \delta_{23})\notag \\
    &= -\nuF \left\{4\Vtriplet(\pi) \left[ \Vsinglet(\pi) + \Vtriplet(\pi) \right] + (-1)^{\ell + 1} \left[ \Vsinglet(\pi) - \Vtriplet(\pi) \right]^2\right\} \frac{\ln2}{2\ell+1} \frac{\Lambda}{\kF} \left[\delta_{13}\delta_{24} + (-1)^{\ell+1} \delta_{14} \delta_{23} \right]\notag \\
    &\quad- \nuF \left( \Vsinglet_{\ell} \right)^2 (\delta_{13} \delta_{24} - \delta_{14} \delta_{23})
    - \nuF \left( \Vtriplet_{\ell} \right)^2 (\delta_{13} \delta_{24} + \delta_{14} \delta_{23})\,.
\end{align}
The beta function from the BCS diagram in the second line on the right hand side of the equation does not mix the singlet and triplet couplings; the RG evolution of the singlet coupling is governed solely by the singlet BCS beta function, and the same is true of the triplet coupling.
Meanwhile, the beta function from the ZS and ZS' diagrams mixes the RG evolution of the singlet and the triplet couplings and contributes to the RG evolution of the singlet or triplet couplings depending on the parity of $\ell$.
When $\ell$ is even, this contributes only to the singlet coupling and when $\ell$ is odd, this contributes only to the triplet coupling.

\bibliography{bibKL}

\begin{thebibliography}{60}%
\makeatletter
\providecommand \@ifxundefined [1]{%
 \@ifx{#1\undefined}
}%
\providecommand \@ifnum [1]{%
 \ifnum #1\expandafter \@firstoftwo
 \else \expandafter \@secondoftwo
 \fi
}%
\providecommand \@ifx [1]{%
 \ifx #1\expandafter \@firstoftwo
 \else \expandafter \@secondoftwo
 \fi
}%
\providecommand \natexlab [1]{#1}%
\providecommand \enquote  [1]{``#1''}%
\providecommand \bibnamefont  [1]{#1}%
\providecommand \bibfnamefont [1]{#1}%
\providecommand \citenamefont [1]{#1}%
\providecommand \href@noop [0]{\@secondoftwo}%
\providecommand \href [0]{\begingroup \@sanitize@url \@href}%
\providecommand \@href[1]{\@@startlink{#1}\@@href}%
\providecommand \@@href[1]{\endgroup#1\@@endlink}%
\providecommand \@sanitize@url [0]{\catcode `\\12\catcode `\$12\catcode
  `\&12\catcode `\#12\catcode `\^12\catcode `\_12\catcode `\%12\relax}%
\providecommand \@@startlink[1]{}%
\providecommand \@@endlink[0]{}%
\providecommand \url  [0]{\begingroup\@sanitize@url \@url }%
\providecommand \@url [1]{\endgroup\@href {#1}{\urlprefix }}%
\providecommand \urlprefix  [0]{URL }%
\providecommand \Eprint [0]{\href }%
\providecommand \doibase [0]{https://doi.org/}%
\providecommand \selectlanguage [0]{\@gobble}%
\providecommand \bibinfo  [0]{\@secondoftwo}%
\providecommand \bibfield  [0]{\@secondoftwo}%
\providecommand \translation [1]{[#1]}%
\providecommand \BibitemOpen [0]{}%
\providecommand \bibitemStop [0]{}%
\providecommand \bibitemNoStop [0]{.\EOS\space}%
\providecommand \EOS [0]{\spacefactor3000\relax}%
\providecommand \BibitemShut  [1]{\csname bibitem#1\endcsname}%
\let\auto@bib@innerbib\@empty
\bibitem [{\citenamefont {Kohn}\ and\ \citenamefont
  {Luttinger}(1965)}]{Kohn:1965zz}%
  \BibitemOpen
  \bibfield  {author} {\bibinfo {author} {\bibfnamefont {W.}~\bibnamefont
  {Kohn}}\ and\ \bibinfo {author} {\bibfnamefont {J.~M.}\ \bibnamefont
  {Luttinger}},\ }\bibfield  {title} {\bibinfo {title} {{New Mechanism for
  Superconductivity}},\ }\href {https://doi.org/10.1103/PhysRevLett.15.524}
  {\bibfield  {journal} {\bibinfo  {journal} {Phys. Rev. Lett.}\ }\textbf
  {\bibinfo {volume} {15}},\ \bibinfo {pages} {524} (\bibinfo {year}
  {1965})}\BibitemShut {NoStop}%
\bibitem [{\citenamefont {Luttinger}(1966)}]{PhysRev.150.202}%
  \BibitemOpen
  \bibfield  {author} {\bibinfo {author} {\bibfnamefont {J.~M.}\ \bibnamefont
  {Luttinger}},\ }\bibfield  {title} {\bibinfo {title} {New mechanism for
  superconductivity},\ }\href {https://doi.org/10.1103/PhysRev.150.202}
  {\bibfield  {journal} {\bibinfo  {journal} {Phys. Rev.}\ }\textbf {\bibinfo
  {volume} {150}},\ \bibinfo {pages} {202} (\bibinfo {year}
  {1966})}\BibitemShut {NoStop}%
\bibitem [{\citenamefont {{Baranov}}\ \emph {et~al.}(1992)\citenamefont
  {{Baranov}}, \citenamefont {{Chubukov}},\ and\ \citenamefont
  {{Kagan}}}]{1992IJMPB...6.2471B}%
  \BibitemOpen
  \bibfield  {author} {\bibinfo {author} {\bibfnamefont {M.~A.}\ \bibnamefont
  {{Baranov}}}, \bibinfo {author} {\bibfnamefont {A.~V.}\ \bibnamefont
  {{Chubukov}}},\ and\ \bibinfo {author} {\bibfnamefont {M.~Y.}\ \bibnamefont
  {{Kagan}}},\ }\bibfield  {title} {\bibinfo {title} {{Superconductivity and
  Superfluidity in Fermi Systems with Repulsive Interactions}},\ }\href
  {https://doi.org/10.1142/S0217979292001249} {\bibfield  {journal} {\bibinfo
  {journal} {International Journal of Modern Physics B}\ }\textbf {\bibinfo
  {volume} {6}},\ \bibinfo {pages} {2471} (\bibinfo {year} {1992})}\BibitemShut
  {NoStop}%
\bibitem [{\citenamefont {{Kagan}}(2013)}]{2013LNP...874.....K}%
  \BibitemOpen
  \bibfield  {author} {\bibinfo {author} {\bibfnamefont {M.~Y.}\ \bibnamefont
  {{Kagan}}},\ }\href {https://doi.org/10.1007/978-94-007-6961-8} {\emph
  {\bibinfo {title} {{Modern Trends in Superconductivity and
  Superfluidity}}}},\ Vol.\ \bibinfo {volume} {874}\ (\bibinfo {year}
  {2013})\BibitemShut {NoStop}%
\bibitem [{\citenamefont {{Maiti}}\ and\ \citenamefont
  {{Chubukov}}(2013)}]{2013AIPC.1550....3M}%
  \BibitemOpen
  \bibfield  {author} {\bibinfo {author} {\bibfnamefont {S.}~\bibnamefont
  {{Maiti}}}\ and\ \bibinfo {author} {\bibfnamefont {A.~V.}\ \bibnamefont
  {{Chubukov}}},\ }\bibfield  {title} {\bibinfo {title} {{Superconductivity
  from repulsive interaction}},\ }in\ \href {https://doi.org/10.1063/1.4818400}
  {\emph {\bibinfo {booktitle} {Lectures on the Physics of Strongly Correlated
  Systems XVII: Seventeenth Training Course in the Physics of Strongly
  Correlated Systems}}},\ \bibinfo {series} {American Institute of Physics
  Conference Series}, Vol.\ \bibinfo {volume} {1550},\ \bibinfo {editor}
  {edited by\ \bibinfo {editor} {\bibfnamefont {A.}~\bibnamefont {{Avella}}}\
  and\ \bibinfo {editor} {\bibfnamefont {F.}~\bibnamefont {{Mancini}}}}\
  (\bibinfo  {publisher} {AIP},\ \bibinfo {year} {2013})\ pp.\ \bibinfo {pages}
  {3--73},\ \Eprint {https://arxiv.org/abs/1305.4609} {arXiv:1305.4609
  [cond-mat.supr-con]} \BibitemShut {NoStop}%
\bibitem [{\citenamefont {Fay}\ and\ \citenamefont
  {Layzer}(1968)}]{PhysRevLett.20.187}%
  \BibitemOpen
  \bibfield  {author} {\bibinfo {author} {\bibfnamefont {D.}~\bibnamefont
  {Fay}}\ and\ \bibinfo {author} {\bibfnamefont {A.}~\bibnamefont {Layzer}},\
  }\bibfield  {title} {\bibinfo {title} {Superfluidity of low-density fermion
  systems},\ }\href {https://doi.org/10.1103/PhysRevLett.20.187} {\bibfield
  {journal} {\bibinfo  {journal} {Phys. Rev. Lett.}\ }\textbf {\bibinfo
  {volume} {20}},\ \bibinfo {pages} {187} (\bibinfo {year} {1968})}\BibitemShut
  {NoStop}%
\bibitem [{\citenamefont {{Kagan}}\ and\ \citenamefont
  {{Chubukov}}(1988)}]{1988JETPL..47..614K}%
  \BibitemOpen
  \bibfield  {author} {\bibinfo {author} {\bibfnamefont {M.~Y.}\ \bibnamefont
  {{Kagan}}}\ and\ \bibinfo {author} {\bibfnamefont {A.~V.}\ \bibnamefont
  {{Chubukov}}},\ }\bibfield  {title} {\bibinfo {title} {{Possibility of a
  superfluid transition in a slightly nonideal Fermi gas with repulsion}},\
  }\href@noop {} {\bibfield  {journal} {\bibinfo  {journal} {Soviet Journal of
  Experimental and Theoretical Physics Letters}\ }\textbf {\bibinfo {volume}
  {47}},\ \bibinfo {pages} {614} (\bibinfo {year} {1988})}\BibitemShut
  {NoStop}%
\bibitem [{\citenamefont {{Gonz{\'a}lez}}(2008)}]{2008PhRvB..78t5431G}%
  \BibitemOpen
  \bibfield  {author} {\bibinfo {author} {\bibfnamefont {J.}~\bibnamefont
  {{Gonz{\'a}lez}}},\ }\bibfield  {title} {\bibinfo {title} {{Kohn-Luttinger
  superconductivity in graphene}},\ }\href
  {https://doi.org/10.1103/PhysRevB.78.205431} {\bibfield  {journal} {\bibinfo
  {journal} {\prb}\ }\textbf {\bibinfo {volume} {78}},\ \bibinfo {eid} {205431}
  (\bibinfo {year} {2008})},\ \Eprint {https://arxiv.org/abs/0807.3914}
  {arXiv:0807.3914 [cond-mat.mes-hall]} \BibitemShut {NoStop}%
\bibitem [{\citenamefont {{Nandkishore}}\ \emph {et~al.}(2014)\citenamefont
  {{Nandkishore}}, \citenamefont {{Thomale}},\ and\ \citenamefont
  {{Chubukov}}}]{2014PhRvB..89n4501N}%
  \BibitemOpen
  \bibfield  {author} {\bibinfo {author} {\bibfnamefont {R.}~\bibnamefont
  {{Nandkishore}}}, \bibinfo {author} {\bibfnamefont {R.}~\bibnamefont
  {{Thomale}}},\ and\ \bibinfo {author} {\bibfnamefont {A.~V.}\ \bibnamefont
  {{Chubukov}}},\ }\bibfield  {title} {\bibinfo {title} {{Superconductivity
  from weak repulsion in hexagonal lattice systems}},\ }\href
  {https://doi.org/10.1103/PhysRevB.89.144501} {\bibfield  {journal} {\bibinfo
  {journal} {\prb}\ }\textbf {\bibinfo {volume} {89}},\ \bibinfo {eid} {144501}
  (\bibinfo {year} {2014})},\ \Eprint {https://arxiv.org/abs/1401.5485}
  {arXiv:1401.5485 [cond-mat.supr-con]} \BibitemShut {NoStop}%
\bibitem [{\citenamefont {{Kagan}}\ \emph
  {et~al.}(2014{\natexlab{a}})\citenamefont {{Kagan}}, \citenamefont
  {{Val'kov}}, \citenamefont {{Mitskan}},\ and\ \citenamefont
  {{Korovushkin}}}]{2014SSCom.188...61K}%
  \BibitemOpen
  \bibfield  {author} {\bibinfo {author} {\bibfnamefont {M.~Y.}\ \bibnamefont
  {{Kagan}}}, \bibinfo {author} {\bibfnamefont {V.~V.}\ \bibnamefont
  {{Val'kov}}}, \bibinfo {author} {\bibfnamefont {V.~A.}\ \bibnamefont
  {{Mitskan}}},\ and\ \bibinfo {author} {\bibfnamefont {M.~M.}\ \bibnamefont
  {{Korovushkin}}},\ }\bibfield  {title} {\bibinfo {title} {{The Kohn-Luttinger
  superconductivity in idealized doped graphene}},\ }\href
  {https://doi.org/10.1016/j.ssc.2014.03.001} {\bibfield  {journal} {\bibinfo
  {journal} {Solid State Communications}\ }\textbf {\bibinfo {volume} {188}},\
  \bibinfo {pages} {61} (\bibinfo {year} {2014}{\natexlab{a}})},\ \Eprint
  {https://arxiv.org/abs/1411.3795} {arXiv:1411.3795 [cond-mat.supr-con]}
  \BibitemShut {NoStop}%
\bibitem [{\citenamefont {{Kagan}}\ \emph
  {et~al.}(2014{\natexlab{b}})\citenamefont {{Kagan}}, \citenamefont
  {{Val'kov}}, \citenamefont {{Mitskan}},\ and\ \citenamefont
  {{Korovushkin}}}]{2014JETP..118..995K}%
  \BibitemOpen
  \bibfield  {author} {\bibinfo {author} {\bibfnamefont {M.~Y.}\ \bibnamefont
  {{Kagan}}}, \bibinfo {author} {\bibfnamefont {V.~V.}\ \bibnamefont
  {{Val'kov}}}, \bibinfo {author} {\bibfnamefont {V.~A.}\ \bibnamefont
  {{Mitskan}}},\ and\ \bibinfo {author} {\bibfnamefont {M.~M.}\ \bibnamefont
  {{Korovushkin}}},\ }\bibfield  {title} {\bibinfo {title} {{The Kohn-Luttinger
  effect and anomalous pairing in new superconducting systems and graphene}},\
  }\href {https://doi.org/10.1134/S1063776114060132} {\bibfield  {journal}
  {\bibinfo  {journal} {Soviet Journal of Experimental and Theoretical
  Physics}\ }\textbf {\bibinfo {volume} {118}},\ \bibinfo {pages} {995}
  (\bibinfo {year} {2014}{\natexlab{b}})},\ \Eprint
  {https://arxiv.org/abs/1411.6812} {arXiv:1411.6812 [cond-mat.supr-con]}
  \BibitemShut {NoStop}%
\bibitem [{\citenamefont {{Lin}}\ and\ \citenamefont
  {{Nandkishore}}(2018)}]{2018PhRvB..98u4521L}%
  \BibitemOpen
  \bibfield  {author} {\bibinfo {author} {\bibfnamefont {Y.-P.}\ \bibnamefont
  {{Lin}}}\ and\ \bibinfo {author} {\bibfnamefont {R.~M.}\ \bibnamefont
  {{Nandkishore}}},\ }\bibfield  {title} {\bibinfo {title} {{Kohn-Luttinger
  superconductivity on two orbital honeycomb lattice}},\ }\href
  {https://doi.org/10.1103/PhysRevB.98.214521} {\bibfield  {journal} {\bibinfo
  {journal} {\prb}\ }\textbf {\bibinfo {volume} {98}},\ \bibinfo {eid} {214521}
  (\bibinfo {year} {2018})},\ \Eprint {https://arxiv.org/abs/1808.05270}
  {arXiv:1808.05270 [cond-mat.supr-con]} \BibitemShut {NoStop}%
\bibitem [{\citenamefont {Gonz\'alez}\ and\ \citenamefont
  {Stauber}(2019)}]{PhysRevLett.122.026801}%
  \BibitemOpen
  \bibfield  {author} {\bibinfo {author} {\bibfnamefont {J.}~\bibnamefont
  {Gonz\'alez}}\ and\ \bibinfo {author} {\bibfnamefont {T.}~\bibnamefont
  {Stauber}},\ }\bibfield  {title} {\bibinfo {title} {Kohn-luttinger
  superconductivity in twisted bilayer graphene},\ }\href
  {https://doi.org/10.1103/PhysRevLett.122.026801} {\bibfield  {journal}
  {\bibinfo  {journal} {Phys. Rev. Lett.}\ }\textbf {\bibinfo {volume} {122}},\
  \bibinfo {pages} {026801} (\bibinfo {year} {2019})}\BibitemShut {NoStop}%
\bibitem [{\citenamefont {{You}}\ and\ \citenamefont
  {{Vishwanath}}(2022)}]{2022PhRvB.105m4524Y}%
  \BibitemOpen
  \bibfield  {author} {\bibinfo {author} {\bibfnamefont {Y.-Z.}\ \bibnamefont
  {{You}}}\ and\ \bibinfo {author} {\bibfnamefont {A.}~\bibnamefont
  {{Vishwanath}}},\ }\bibfield  {title} {\bibinfo {title} {{Kohn-Luttinger
  superconductivity and intervalley coherence in rhombohedral trilayer
  graphene}},\ }\href {https://doi.org/10.1103/PhysRevB.105.134524} {\bibfield
  {journal} {\bibinfo  {journal} {\prb}\ }\textbf {\bibinfo {volume} {105}},\
  \bibinfo {eid} {134524} (\bibinfo {year} {2022})}\BibitemShut {NoStop}%
\bibitem [{\citenamefont {{Cea}}\ \emph {et~al.}(2022)\citenamefont {{Cea}},
  \citenamefont {{Pantale{\'o}n}}, \citenamefont {{Phong}},\ and\ \citenamefont
  {{Guinea}}}]{2022PhRvB.105g5432C}%
  \BibitemOpen
  \bibfield  {author} {\bibinfo {author} {\bibfnamefont {T.}~\bibnamefont
  {{Cea}}}, \bibinfo {author} {\bibfnamefont {P.~A.}\ \bibnamefont
  {{Pantale{\'o}n}}}, \bibinfo {author} {\bibfnamefont {V.~T.}\ \bibnamefont
  {{Phong}}},\ and\ \bibinfo {author} {\bibfnamefont {F.}~\bibnamefont
  {{Guinea}}},\ }\bibfield  {title} {\bibinfo {title} {{Superconductivity from
  repulsive interactions in rhombohedral trilayer graphene: A
  Kohn-Luttinger-like mechanism}},\ }\href
  {https://doi.org/10.1103/PhysRevB.105.075432} {\bibfield  {journal} {\bibinfo
   {journal} {\prb}\ }\textbf {\bibinfo {volume} {105}},\ \bibinfo {eid}
  {075432} (\bibinfo {year} {2022})},\ \Eprint
  {https://arxiv.org/abs/2109.04345} {arXiv:2109.04345 [cond-mat.mes-hall]}
  \BibitemShut {NoStop}%
\bibitem [{\citenamefont {{Herrera}}\ \emph {et~al.}(2024)\citenamefont
  {{Herrera}}, \citenamefont {{Parra-Martinez}}, \citenamefont {{Rosenzweig}},
  \citenamefont {{Matta}}, \citenamefont {{Polley}}, \citenamefont {{Kuster}},
  \citenamefont {{Starke}}, \citenamefont {{Guinea}}, \citenamefont
  {{Silva-Guillen}}, \citenamefont {{Naumis}},\ and\ \citenamefont
  {{Pantaleon}}}]{2024arXiv240805271H}%
  \BibitemOpen
  \bibfield  {author} {\bibinfo {author} {\bibfnamefont {S.}~\bibnamefont
  {{Herrera}}}, \bibinfo {author} {\bibfnamefont {G.}~\bibnamefont
  {{Parra-Martinez}}}, \bibinfo {author} {\bibfnamefont {P.}~\bibnamefont
  {{Rosenzweig}}}, \bibinfo {author} {\bibfnamefont {B.}~\bibnamefont
  {{Matta}}}, \bibinfo {author} {\bibfnamefont {C.~M.}\ \bibnamefont
  {{Polley}}}, \bibinfo {author} {\bibfnamefont {K.}~\bibnamefont {{Kuster}}},
  \bibinfo {author} {\bibfnamefont {U.}~\bibnamefont {{Starke}}}, \bibinfo
  {author} {\bibfnamefont {F.}~\bibnamefont {{Guinea}}}, \bibinfo {author}
  {\bibfnamefont {J.~A.}\ \bibnamefont {{Silva-Guillen}}}, \bibinfo {author}
  {\bibfnamefont {G.~G.}\ \bibnamefont {{Naumis}}},\ and\ \bibinfo {author}
  {\bibfnamefont {P.~A.}\ \bibnamefont {{Pantaleon}}},\ }\bibfield  {title}
  {\bibinfo {title} {{Electronic Structure and Kohn-Luttinger Superconductivity
  of Heavily-Doped Single-Layer Graphene}},\ }\href
  {https://doi.org/10.48550/arXiv.2408.05271} {\bibfield  {journal} {\bibinfo
  {journal} {arXiv e-prints}\ ,\ \bibinfo {eid} {arXiv:2408.05271}} (\bibinfo
  {year} {2024})},\ \Eprint {https://arxiv.org/abs/2408.05271}
  {arXiv:2408.05271 [cond-mat.mes-hall]} \BibitemShut {NoStop}%
\bibitem [{\citenamefont {{Qin}}\ \emph {et~al.}(2024)\citenamefont {{Qin}},
  \citenamefont {{Qiu}},\ and\ \citenamefont {{Wu}}}]{2024arXiv240916114Q}%
  \BibitemOpen
  \bibfield  {author} {\bibinfo {author} {\bibfnamefont {W.}~\bibnamefont
  {{Qin}}}, \bibinfo {author} {\bibfnamefont {W.-X.}\ \bibnamefont {{Qiu}}},\
  and\ \bibinfo {author} {\bibfnamefont {F.}~\bibnamefont {{Wu}}},\ }\bibfield
  {title} {\bibinfo {title} {{Kohn-Luttinger Mechanism of Superconductivity in
  Twisted Bilayer WSe$_2$: Gate-Tunable Unconventional Pairing Symmetry}},\
  }\href {https://doi.org/10.48550/arXiv.2409.16114} {\bibfield  {journal}
  {\bibinfo  {journal} {arXiv e-prints}\ ,\ \bibinfo {eid} {arXiv:2409.16114}}
  (\bibinfo {year} {2024})},\ \Eprint {https://arxiv.org/abs/2409.16114}
  {arXiv:2409.16114 [cond-mat.supr-con]} \BibitemShut {NoStop}%
\bibitem [{\citenamefont {{Jahin}}\ and\ \citenamefont
  {{Lin}}(2024)}]{2024arXiv241109664J}%
  \BibitemOpen
  \bibfield  {author} {\bibinfo {author} {\bibfnamefont {A.}~\bibnamefont
  {{Jahin}}}\ and\ \bibinfo {author} {\bibfnamefont {S.-Z.}\ \bibnamefont
  {{Lin}}},\ }\bibfield  {title} {\bibinfo {title} {{Enhanced Kohn-Luttinger
  topological superconductivity in bands with nontrivial geometry}},\ }\href
  {https://doi.org/10.48550/arXiv.2411.09664} {\bibfield  {journal} {\bibinfo
  {journal} {arXiv e-prints}\ ,\ \bibinfo {eid} {arXiv:2411.09664}} (\bibinfo
  {year} {2024})},\ \Eprint {https://arxiv.org/abs/2411.09664}
  {arXiv:2411.09664 [cond-mat.supr-con]} \BibitemShut {NoStop}%
\bibitem [{\citenamefont {Leggett}(1975)}]{Leggett:1975te}%
  \BibitemOpen
  \bibfield  {author} {\bibinfo {author} {\bibfnamefont {A.~J.}\ \bibnamefont
  {Leggett}},\ }\bibfield  {title} {\bibinfo {title} {{A theoretical
  description of the new phases of liquid He-3}},\ }\href
  {https://doi.org/10.1103/RevModPhys.47.331} {\bibfield  {journal} {\bibinfo
  {journal} {Rev. Mod. Phys.}\ }\textbf {\bibinfo {volume} {47}},\ \bibinfo
  {pages} {331} (\bibinfo {year} {1975})},\ \bibinfo {note} {[Erratum:
  Rev.Mod.Phys. 48, 357--357 (1976)]}\BibitemShut {NoStop}%
\bibitem [{\citenamefont {Lee}(1997)}]{Lee:1997zzh}%
  \BibitemOpen
  \bibfield  {author} {\bibinfo {author} {\bibfnamefont {D.~M.}\ \bibnamefont
  {Lee}},\ }\bibfield  {title} {\bibinfo {title} {{The extraordinary phases of
  liquid He-3}},\ }\href {https://doi.org/10.1103/RevModPhys.69.645} {\bibfield
   {journal} {\bibinfo  {journal} {Rev. Mod. Phys.}\ }\textbf {\bibinfo
  {volume} {69}},\ \bibinfo {pages} {645} (\bibinfo {year} {1997})},\ \bibinfo
  {note} {[Erratum: Rev.Mod.Phys. 70, 319--319 (1998)]}\BibitemShut {NoStop}%
\bibitem [{\citenamefont {{Sauls}}(1994)}]{1994AdPhy..43..113S}%
  \BibitemOpen
  \bibfield  {author} {\bibinfo {author} {\bibfnamefont {J.~A.}\ \bibnamefont
  {{Sauls}}},\ }\bibfield  {title} {\bibinfo {title} {{The order parameter for
  the superconducting phases of UPt3}},\ }\href
  {https://doi.org/10.1080/00018739400101475} {\bibfield  {journal} {\bibinfo
  {journal} {Advances in Physics}\ }\textbf {\bibinfo {volume} {43}},\ \bibinfo
  {pages} {113} (\bibinfo {year} {1994})},\ \Eprint
  {https://arxiv.org/abs/1812.09984} {arXiv:1812.09984 [cond-mat.supr-con]}
  \BibitemShut {NoStop}%
\bibitem [{\citenamefont {Baranov}(2008)}]{BARANOV200871}%
  \BibitemOpen
  \bibfield  {author} {\bibinfo {author} {\bibfnamefont {M.}~\bibnamefont
  {Baranov}},\ }\bibfield  {title} {\bibinfo {title} {Theoretical progress in
  many-body physics with ultracold dipolar gases},\ }\href
  {https://doi.org/https://doi.org/10.1016/j.physrep.2008.04.007} {\bibfield
  {journal} {\bibinfo  {journal} {Physics Reports}\ }\textbf {\bibinfo {volume}
  {464}},\ \bibinfo {pages} {71} (\bibinfo {year} {2008})}\BibitemShut
  {NoStop}%
\bibitem [{\citenamefont {{Kanoda}}\ and\ \citenamefont
  {{Kato}}(2011)}]{2011ARCMP...2..167K}%
  \BibitemOpen
  \bibfield  {author} {\bibinfo {author} {\bibfnamefont {K.}~\bibnamefont
  {{Kanoda}}}\ and\ \bibinfo {author} {\bibfnamefont {R.}~\bibnamefont
  {{Kato}}},\ }\bibfield  {title} {\bibinfo {title} {{Mott Physics in Organic
  Conductors with Triangular Lattices}},\ }\href
  {https://doi.org/10.1146/annurev-conmatphys-062910-140521} {\bibfield
  {journal} {\bibinfo  {journal} {Annual Review of Condensed Matter Physics}\
  }\textbf {\bibinfo {volume} {2}},\ \bibinfo {pages} {167} (\bibinfo {year}
  {2011})}\BibitemShut {NoStop}%
\bibitem [{\citenamefont {{Kallin}}(2012)}]{2012RPPh...75d2501K}%
  \BibitemOpen
  \bibfield  {author} {\bibinfo {author} {\bibfnamefont {C.}~\bibnamefont
  {{Kallin}}},\ }\bibfield  {title} {\bibinfo {title} {{Chiral p-wave order in
  Sr$_{2}$RuO$_{4}$}},\ }\href {https://doi.org/10.1088/0034-4885/75/4/042501}
  {\bibfield  {journal} {\bibinfo  {journal} {Reports on Progress in Physics}\
  }\textbf {\bibinfo {volume} {75}},\ \bibinfo {eid} {042501} (\bibinfo {year}
  {2012})},\ \Eprint {https://arxiv.org/abs/1210.2992} {arXiv:1210.2992
  [cond-mat.supr-con]} \BibitemShut {NoStop}%
\bibitem [{\citenamefont {Hoffberg}\ \emph {et~al.}(1970)\citenamefont
  {Hoffberg}, \citenamefont {Glassgold}, \citenamefont {Richardson},\ and\
  \citenamefont {Ruderman}}]{Hoffberg:1970vqj}%
  \BibitemOpen
  \bibfield  {author} {\bibinfo {author} {\bibfnamefont {M.}~\bibnamefont
  {Hoffberg}}, \bibinfo {author} {\bibfnamefont {A.~E.}\ \bibnamefont
  {Glassgold}}, \bibinfo {author} {\bibfnamefont {R.~W.}\ \bibnamefont
  {Richardson}},\ and\ \bibinfo {author} {\bibfnamefont {M.}~\bibnamefont
  {Ruderman}},\ }\bibfield  {title} {\bibinfo {title} {{Anisotropic
  Superfluidity in Neutron Star Matter}},\ }\href
  {https://doi.org/10.1103/PhysRevLett.24.775} {\bibfield  {journal} {\bibinfo
  {journal} {Phys. Rev. Lett.}\ }\textbf {\bibinfo {volume} {24}},\ \bibinfo
  {pages} {775} (\bibinfo {year} {1970})}\BibitemShut {NoStop}%
\bibitem [{\citenamefont {Tamagaki}(1970)}]{Tamagaki:1970ptp}%
  \BibitemOpen
  \bibfield  {author} {\bibinfo {author} {\bibfnamefont {R.}~\bibnamefont
  {Tamagaki}},\ }\bibfield  {title} {\bibinfo {title} {{Superfluid State in
  Neutron Star Matter. I. Generalized Bogoliubov Transformation and Existence
  of ${}^3P_2$ Gap at High Density}},\ }\href
  {https://doi.org/10.1143/PTP.44.905} {\bibfield  {journal} {\bibinfo
  {journal} {Prog. Theor. Phys.}\ }\textbf {\bibinfo {volume} {44}},\ \bibinfo
  {pages} {905} (\bibinfo {year} {1970})}\BibitemShut {NoStop}%
\bibitem [{\citenamefont {Schwenk}\ and\ \citenamefont
  {Friman}(2004)}]{Schwenk:2003bc}%
  \BibitemOpen
  \bibfield  {author} {\bibinfo {author} {\bibfnamefont {A.}~\bibnamefont
  {Schwenk}}\ and\ \bibinfo {author} {\bibfnamefont {B.}~\bibnamefont
  {Friman}},\ }\bibfield  {title} {\bibinfo {title} {{Polarization
  contributions to the spin dependence of the effective interaction in neutron
  matter}},\ }\href {https://doi.org/10.1103/PhysRevLett.92.082501} {\bibfield
  {journal} {\bibinfo  {journal} {Phys. Rev. Lett.}\ }\textbf {\bibinfo
  {volume} {92}},\ \bibinfo {pages} {082501} (\bibinfo {year} {2004})},\
  \Eprint {https://arxiv.org/abs/nucl-th/0307089} {arXiv:nucl-th/0307089}
  \BibitemShut {NoStop}%
\bibitem [{\citenamefont {Kumamoto}\ and\ \citenamefont
  {Reddy}(2024)}]{Kumamoto:2024muw}%
  \BibitemOpen
  \bibfield  {author} {\bibinfo {author} {\bibfnamefont {M.}~\bibnamefont
  {Kumamoto}}\ and\ \bibinfo {author} {\bibfnamefont {S.}~\bibnamefont
  {Reddy}},\ }\bibfield  {title} {\bibinfo {title} {{Kohn-Luttinger effect in
  dense matter and its implications for neutron stars}},\ }\href
  {https://doi.org/10.1103/PhysRevC.110.025804} {\bibfield  {journal} {\bibinfo
   {journal} {Phys. Rev. C}\ }\textbf {\bibinfo {volume} {110}},\ \bibinfo
  {pages} {025804} (\bibinfo {year} {2024})},\ \Eprint
  {https://arxiv.org/abs/2405.12243} {arXiv:2405.12243 [nucl-th]} \BibitemShut
  {NoStop}%
\bibitem [{\citenamefont {Page}\ \emph {et~al.}(2011)\citenamefont {Page},
  \citenamefont {Prakash}, \citenamefont {Lattimer},\ and\ \citenamefont
  {Steiner}}]{Page:2010aw}%
  \BibitemOpen
  \bibfield  {author} {\bibinfo {author} {\bibfnamefont {D.}~\bibnamefont
  {Page}}, \bibinfo {author} {\bibfnamefont {M.}~\bibnamefont {Prakash}},
  \bibinfo {author} {\bibfnamefont {J.~M.}\ \bibnamefont {Lattimer}},\ and\
  \bibinfo {author} {\bibfnamefont {A.~W.}\ \bibnamefont {Steiner}},\
  }\bibfield  {title} {\bibinfo {title} {{Rapid Cooling of the Neutron Star in
  Cassiopeia A Triggered by Neutron Superfluidity in Dense Matter}},\ }\href
  {https://doi.org/10.1103/PhysRevLett.106.081101} {\bibfield  {journal}
  {\bibinfo  {journal} {Phys. Rev. Lett.}\ }\textbf {\bibinfo {volume} {106}},\
  \bibinfo {pages} {081101} (\bibinfo {year} {2011})},\ \Eprint
  {https://arxiv.org/abs/1011.6142} {arXiv:1011.6142 [astro-ph.HE]}
  \BibitemShut {NoStop}%
\bibitem [{\citenamefont {Shternin}\ \emph {et~al.}(2011)\citenamefont
  {Shternin}, \citenamefont {Yakovlev}, \citenamefont {Heinke}, \citenamefont
  {Ho},\ and\ \citenamefont {Patnaude}}]{Shternin:2010qi}%
  \BibitemOpen
  \bibfield  {author} {\bibinfo {author} {\bibfnamefont {P.~S.}\ \bibnamefont
  {Shternin}}, \bibinfo {author} {\bibfnamefont {D.~G.}\ \bibnamefont
  {Yakovlev}}, \bibinfo {author} {\bibfnamefont {C.~O.}\ \bibnamefont
  {Heinke}}, \bibinfo {author} {\bibfnamefont {W.~C.~G.}\ \bibnamefont {Ho}},\
  and\ \bibinfo {author} {\bibfnamefont {D.~J.}\ \bibnamefont {Patnaude}},\
  }\bibfield  {title} {\bibinfo {title} {{Cooling neutron star in the
  Cassiopeia\textasciitilde{}A supernova remnant: Evidence for superfluidity in
  the core}},\ }\href {https://doi.org/10.1111/j.1745-3933.2011.01015.x}
  {\bibfield  {journal} {\bibinfo  {journal} {Mon. Not. Roy. Astron. Soc.}\
  }\textbf {\bibinfo {volume} {412}},\ \bibinfo {pages} {L108} (\bibinfo {year}
  {2011})},\ \Eprint {https://arxiv.org/abs/1012.0045} {arXiv:1012.0045
  [astro-ph.SR]} \BibitemShut {NoStop}%
\bibitem [{\citenamefont {Sch\"afer}(2006)}]{Schafer:2006ue}%
  \BibitemOpen
  \bibfield  {author} {\bibinfo {author} {\bibfnamefont {T.}~\bibnamefont
  {Sch\"afer}},\ }\bibfield  {title} {\bibinfo {title} {{The Kohn-Luttinger
  effect in gauge theories}},\ }\href
  {https://doi.org/10.1103/PhysRevD.74.054009} {\bibfield  {journal} {\bibinfo
  {journal} {Phys. Rev. D}\ }\textbf {\bibinfo {volume} {74}},\ \bibinfo
  {pages} {054009} (\bibinfo {year} {2006})},\ \Eprint
  {https://arxiv.org/abs/hep-ph/0606026} {arXiv:hep-ph/0606026} \BibitemShut
  {NoStop}%
\bibitem [{\citenamefont {{Efremov}}\ \emph {et~al.}(2000)\citenamefont
  {{Efremov}}, \citenamefont {{Mar'Enko}}, \citenamefont {{Baranov}},\ and\
  \citenamefont {{Kagan}}}]{2000JETP...90..861E}%
  \BibitemOpen
  \bibfield  {author} {\bibinfo {author} {\bibfnamefont {D.~V.}\ \bibnamefont
  {{Efremov}}}, \bibinfo {author} {\bibfnamefont {M.~S.}\ \bibnamefont
  {{Mar'Enko}}}, \bibinfo {author} {\bibfnamefont {M.~A.}\ \bibnamefont
  {{Baranov}}},\ and\ \bibinfo {author} {\bibfnamefont {M.~Y.}\ \bibnamefont
  {{Kagan}}},\ }\bibfield  {title} {\bibinfo {title} {{Superfluid Transition
  Temperature in a Fermi Gas with Repulsion Allowing for Higher Orders of
  Perturbation Theory}},\ }\href {https://doi.org/10.1134/1.559173} {\bibfield
  {journal} {\bibinfo  {journal} {Soviet Journal of Experimental and
  Theoretical Physics}\ }\textbf {\bibinfo {volume} {90}},\ \bibinfo {pages}
  {861} (\bibinfo {year} {2000})},\ \Eprint
  {https://arxiv.org/abs/cond-mat/0007334} {arXiv:cond-mat/0007334
  [cond-mat.supr-con]} \BibitemShut {NoStop}%
\bibitem [{\citenamefont {Benfatto}\ and\ \citenamefont
  {Gallavotti}(1990{\natexlab{a}})}]{BenfattoGallavotti1990}%
  \BibitemOpen
  \bibfield  {author} {\bibinfo {author} {\bibfnamefont {G.}~\bibnamefont
  {Benfatto}}\ and\ \bibinfo {author} {\bibfnamefont {G.}~\bibnamefont
  {Gallavotti}},\ }\bibfield  {title} {\bibinfo {title} {Perturbation theory of
  the fermi surface in a quantum liquid. a general quasiparticle formalism and
  one-dimensional systems},\ }\href {https://doi.org/10.1007/BF01025844}
  {\bibfield  {journal} {\bibinfo  {journal} {Journal of Statistical Physics}\
  }\textbf {\bibinfo {volume} {59}},\ \bibinfo {pages} {541} (\bibinfo {year}
  {1990}{\natexlab{a}})}\BibitemShut {NoStop}%
\bibitem [{\citenamefont {Benfatto}\ and\ \citenamefont
  {Gallavotti}(1990{\natexlab{b}})}]{Benfatto:1990zz}%
  \BibitemOpen
  \bibfield  {author} {\bibinfo {author} {\bibfnamefont {G.}~\bibnamefont
  {Benfatto}}\ and\ \bibinfo {author} {\bibfnamefont {G.}~\bibnamefont
  {Gallavotti}},\ }\bibfield  {title} {\bibinfo {title} {{Renormalization-group
  approach to the theory of the Fermi surface}},\ }\href
  {https://doi.org/10.1103/PhysRevB.42.9967} {\bibfield  {journal} {\bibinfo
  {journal} {Phys. Rev. B}\ }\textbf {\bibinfo {volume} {42}},\ \bibinfo
  {pages} {9967} (\bibinfo {year} {1990}{\natexlab{b}})}\BibitemShut {NoStop}%
\bibitem [{\citenamefont {Benfatto}\ and\ \citenamefont
  {Gallavotti}(1996)}]{Benfatto:1996ng}%
  \BibitemOpen
  \bibfield  {author} {\bibinfo {author} {\bibfnamefont {G.}~\bibnamefont
  {Benfatto}}\ and\ \bibinfo {author} {\bibfnamefont {G.}~\bibnamefont
  {Gallavotti}},\ }\href@noop {} {\emph {\bibinfo {title} {{Renormalization
  group}}}}\ (\bibinfo  {publisher} {Princeton University Press},\ \bibinfo
  {year} {1996})\BibitemShut {NoStop}%
\bibitem [{\citenamefont {Feldman}\ and\ \citenamefont
  {Trubowitz}(1990)}]{Feldman1990PerturbationTF}%
  \BibitemOpen
  \bibfield  {author} {\bibinfo {author} {\bibfnamefont {J.~J.}\ \bibnamefont
  {Feldman}}\ and\ \bibinfo {author} {\bibfnamefont {E.}~\bibnamefont
  {Trubowitz}},\ }\bibfield  {title} {\bibinfo {title} {Perturbation theory for
  many fermion systems},\ }\href {https://doi.org/10.5169/seals-116219}
  {\bibfield  {journal} {\bibinfo  {journal} {Helvetica Physica Acta}\ }\textbf
  {\bibinfo {volume} {63}},\ \bibinfo {pages} {156} (\bibinfo {year}
  {1990})}\BibitemShut {NoStop}%
\bibitem [{\citenamefont {Feldman}\ and\ \citenamefont
  {Trubowitz}(1991)}]{Feldman1991TheFO}%
  \BibitemOpen
  \bibfield  {author} {\bibinfo {author} {\bibfnamefont {J.~J.}\ \bibnamefont
  {Feldman}}\ and\ \bibinfo {author} {\bibfnamefont {E.}~\bibnamefont
  {Trubowitz}},\ }\bibfield  {title} {\bibinfo {title} {The flow of an
  electron-phonon system to the superconducing state},\ }\href
  {https://doi.org/10.5169/seals-116309} {\bibfield  {journal} {\bibinfo
  {journal} {Helvetica Physica Acta}\ }\textbf {\bibinfo {volume} {64}},\
  \bibinfo {pages} {213} (\bibinfo {year} {1991})}\BibitemShut {NoStop}%
\bibitem [{\citenamefont {Feldman}\ \emph {et~al.}(1992)\citenamefont
  {Feldman}, \citenamefont {Magnen}, \citenamefont {Rivasseau},\ and\
  \citenamefont {Trubowitz}}]{Feldman1992AnIV}%
  \BibitemOpen
  \bibfield  {author} {\bibinfo {author} {\bibfnamefont {J.~J.}\ \bibnamefont
  {Feldman}}, \bibinfo {author} {\bibfnamefont {J.~L.}\ \bibnamefont {Magnen}},
  \bibinfo {author} {\bibfnamefont {V.}~\bibnamefont {Rivasseau}},\ and\
  \bibinfo {author} {\bibfnamefont {E.}~\bibnamefont {Trubowitz}},\ }\bibfield
  {title} {\bibinfo {title} {An infinite volume expansion for many fermion
  green's functions},\ }\href {https://doi.org/10.5169/seals-116509} {\bibfield
   {journal} {\bibinfo  {journal} {Helvetica Physica Acta}\ }\textbf {\bibinfo
  {volume} {65}},\ \bibinfo {pages} {679} (\bibinfo {year} {1992})}\BibitemShut
  {NoStop}%
\bibitem [{\citenamefont {Feldman}\ \emph {et~al.}(1993)\citenamefont
  {Feldman}, \citenamefont {Magnen}, \citenamefont {Rivasseau},\ and\
  \citenamefont {Trubowitz}}]{Feldman:1993ck}%
  \BibitemOpen
  \bibfield  {author} {\bibinfo {author} {\bibfnamefont {J.}~\bibnamefont
  {Feldman}}, \bibinfo {author} {\bibfnamefont {J.}~\bibnamefont {Magnen}},
  \bibinfo {author} {\bibfnamefont {V.}~\bibnamefont {Rivasseau}},\ and\
  \bibinfo {author} {\bibfnamefont {E.}~\bibnamefont {Trubowitz}},\ }\bibfield
  {title} {\bibinfo {title} {{An Intrinsic 1/N expansion for many fermion
  systems}},\ }\href {https://doi.org/10.1209/0295-5075/24/6/002} {\bibfield
  {journal} {\bibinfo  {journal} {EPL}\ }\textbf {\bibinfo {volume} {24}},\
  \bibinfo {pages} {437} (\bibinfo {year} {1993})}\BibitemShut {NoStop}%
\bibitem [{\citenamefont {Shankar}(1991)}]{SHANKAR1991530}%
  \BibitemOpen
  \bibfield  {author} {\bibinfo {author} {\bibfnamefont {R.}~\bibnamefont
  {Shankar}},\ }\bibfield  {title} {\bibinfo {title} {Renormalization group for
  interacting fermions in $d > 1$},\ }\href
  {https://doi.org/https://doi.org/10.1016/0378-4371(91)90197-K} {\bibfield
  {journal} {\bibinfo  {journal} {Physica A: Statistical Mechanics and its
  Applications}\ }\textbf {\bibinfo {volume} {177}},\ \bibinfo {pages} {530}
  (\bibinfo {year} {1991})}\BibitemShut {NoStop}%
\bibitem [{\citenamefont {Shankar}(1994)}]{Shankar:1993pf}%
  \BibitemOpen
  \bibfield  {author} {\bibinfo {author} {\bibfnamefont {R.}~\bibnamefont
  {Shankar}},\ }\bibfield  {title} {\bibinfo {title} {{Renormalization group
  approach to interacting fermions}},\ }\href
  {https://doi.org/10.1103/RevModPhys.66.129} {\bibfield  {journal} {\bibinfo
  {journal} {Rev. Mod. Phys.}\ }\textbf {\bibinfo {volume} {66}},\ \bibinfo
  {pages} {129} (\bibinfo {year} {1994})},\ \Eprint
  {https://arxiv.org/abs/cond-mat/9307009} {arXiv:cond-mat/9307009}
  \BibitemShut {NoStop}%
\bibitem [{\citenamefont {Polchinski}(1992)}]{Polchinski:1992ed}%
  \BibitemOpen
  \bibfield  {author} {\bibinfo {author} {\bibfnamefont {J.}~\bibnamefont
  {Polchinski}},\ }\bibfield  {title} {\bibinfo {title} {{Effective field
  theory and the Fermi surface}},\ }in\ \href@noop {} {\emph {\bibinfo
  {booktitle} {{Theoretical Advanced Study Institute (TASI 92): From Black
  Holes and Strings to Particles}}}}\ (\bibinfo {year} {1992})\ pp.\ \bibinfo
  {pages} {0235--276},\ \Eprint {https://arxiv.org/abs/hep-th/9210046}
  {arXiv:hep-th/9210046} \BibitemShut {NoStop}%
\bibitem [{\citenamefont {{Anderson}}\ and\ \citenamefont
  {{Yuval}}(1969)}]{1969PhRvL..23...89A}%
  \BibitemOpen
  \bibfield  {author} {\bibinfo {author} {\bibfnamefont {P.~W.}\ \bibnamefont
  {{Anderson}}}\ and\ \bibinfo {author} {\bibfnamefont {G.}~\bibnamefont
  {{Yuval}}},\ }\bibfield  {title} {\bibinfo {title} {{Exact Results in the
  Kondo Problem: Equivalence to a Classical One-Dimensional Coulomb Gas}},\
  }\href {https://doi.org/10.1103/PhysRevLett.23.89} {\bibfield  {journal}
  {\bibinfo  {journal} {\prl}\ }\textbf {\bibinfo {volume} {23}},\ \bibinfo
  {pages} {89} (\bibinfo {year} {1969})}\BibitemShut {NoStop}%
\bibitem [{\citenamefont {{Nozi{\`e}res}}(1974)}]{1974JLTP...17...31N}%
  \BibitemOpen
  \bibfield  {author} {\bibinfo {author} {\bibfnamefont {P.}~\bibnamefont
  {{Nozi{\`e}res}}},\ }\bibfield  {title} {\bibinfo {title} {{A
  ``fermi-liquid'' description of the Kondo problem at low temperatures}},\
  }\href {https://doi.org/10.1007/BF00654541} {\bibfield  {journal} {\bibinfo
  {journal} {Journal of Low Temperature Physics}\ }\textbf {\bibinfo {volume}
  {17}},\ \bibinfo {pages} {31} (\bibinfo {year} {1974})}\BibitemShut {NoStop}%
\bibitem [{\citenamefont {{Wilson}}(1975)}]{1975RvMP...47..773W}%
  \BibitemOpen
  \bibfield  {author} {\bibinfo {author} {\bibfnamefont {K.~G.}\ \bibnamefont
  {{Wilson}}},\ }\bibfield  {title} {\bibinfo {title} {{The renormalization
  group: Critical phenomena and the Kondo problem}},\ }\href
  {https://doi.org/10.1103/RevModPhys.47.773} {\bibfield  {journal} {\bibinfo
  {journal} {Reviews of Modern Physics}\ }\textbf {\bibinfo {volume} {47}},\
  \bibinfo {pages} {773} (\bibinfo {year} {1975})}\BibitemShut {NoStop}%
\bibitem [{\citenamefont {{Krishna-Murthy}}\ \emph
  {et~al.}(1980{\natexlab{a}})\citenamefont {{Krishna-Murthy}}, \citenamefont
  {{Wilkins}},\ and\ \citenamefont {{Wilson}}}]{1980PhRvB..21.1003K}%
  \BibitemOpen
  \bibfield  {author} {\bibinfo {author} {\bibfnamefont {H.~R.}\ \bibnamefont
  {{Krishna-Murthy}}}, \bibinfo {author} {\bibfnamefont {J.~W.}\ \bibnamefont
  {{Wilkins}}},\ and\ \bibinfo {author} {\bibfnamefont {K.~G.}\ \bibnamefont
  {{Wilson}}},\ }\bibfield  {title} {\bibinfo {title} {{Renormalization-group
  approach to the Anderson model of dilute magnetic alloys. I. Static
  properties for the symmetric case}},\ }\href
  {https://doi.org/10.1103/PhysRevB.21.1003} {\bibfield  {journal} {\bibinfo
  {journal} {\prb}\ }\textbf {\bibinfo {volume} {21}},\ \bibinfo {pages} {1003}
  (\bibinfo {year} {1980}{\natexlab{a}})}\BibitemShut {NoStop}%
\bibitem [{\citenamefont {{Krishna-Murthy}}\ \emph
  {et~al.}(1980{\natexlab{b}})\citenamefont {{Krishna-Murthy}}, \citenamefont
  {{Wilkins}},\ and\ \citenamefont {{Wilson}}}]{1980PhRvB..21.1044K}%
  \BibitemOpen
  \bibfield  {author} {\bibinfo {author} {\bibfnamefont {H.~R.}\ \bibnamefont
  {{Krishna-Murthy}}}, \bibinfo {author} {\bibfnamefont {J.~W.}\ \bibnamefont
  {{Wilkins}}},\ and\ \bibinfo {author} {\bibfnamefont {K.~G.}\ \bibnamefont
  {{Wilson}}},\ }\bibfield  {title} {\bibinfo {title} {{Renormalization-group
  approach to the Anderson model of dilute magnetic alloys. II. Static
  properties for the asymmetric case}},\ }\href
  {https://doi.org/10.1103/PhysRevB.21.1044} {\bibfield  {journal} {\bibinfo
  {journal} {\prb}\ }\textbf {\bibinfo {volume} {21}},\ \bibinfo {pages} {1044}
  (\bibinfo {year} {1980}{\natexlab{b}})}\BibitemShut {NoStop}%
\bibitem [{\citenamefont {Schwenk}\ \emph {et~al.}(2002)\citenamefont
  {Schwenk}, \citenamefont {Brown},\ and\ \citenamefont
  {Friman}}]{Schwenk:2001hg}%
  \BibitemOpen
  \bibfield  {author} {\bibinfo {author} {\bibfnamefont {A.}~\bibnamefont
  {Schwenk}}, \bibinfo {author} {\bibfnamefont {G.~E.}\ \bibnamefont {Brown}},\
  and\ \bibinfo {author} {\bibfnamefont {B.}~\bibnamefont {Friman}},\
  }\bibfield  {title} {\bibinfo {title} {{Low momentum nucleon-nucleon
  interaction and Fermi liquid theory}},\ }\href
  {https://doi.org/10.1016/S0375-9474(01)01673-6} {\bibfield  {journal}
  {\bibinfo  {journal} {Nucl. Phys. A}\ }\textbf {\bibinfo {volume} {703}},\
  \bibinfo {pages} {745} (\bibinfo {year} {2002})},\ \Eprint
  {https://arxiv.org/abs/nucl-th/0109059} {arXiv:nucl-th/0109059} \BibitemShut
  {NoStop}%
\bibitem [{\citenamefont {Schwenk}\ \emph {et~al.}(2003)\citenamefont
  {Schwenk}, \citenamefont {Friman},\ and\ \citenamefont
  {Brown}}]{Schwenk:2002fq}%
  \BibitemOpen
  \bibfield  {author} {\bibinfo {author} {\bibfnamefont {A.}~\bibnamefont
  {Schwenk}}, \bibinfo {author} {\bibfnamefont {B.}~\bibnamefont {Friman}},\
  and\ \bibinfo {author} {\bibfnamefont {G.~E.}\ \bibnamefont {Brown}},\
  }\bibfield  {title} {\bibinfo {title} {{Renormalization group approach to
  neutron matter: Quasiparticle interactions, superfluid gaps and the equation
  of state}},\ }\href {https://doi.org/10.1016/S0375-9474(02)01290-3}
  {\bibfield  {journal} {\bibinfo  {journal} {Nucl. Phys. A}\ }\textbf
  {\bibinfo {volume} {713}},\ \bibinfo {pages} {191} (\bibinfo {year}
  {2003})},\ \Eprint {https://arxiv.org/abs/nucl-th/0207004}
  {arXiv:nucl-th/0207004} \BibitemShut {NoStop}%
\bibitem [{\citenamefont {Son}(1999)}]{Son:1998uk}%
  \BibitemOpen
  \bibfield  {author} {\bibinfo {author} {\bibfnamefont {D.~T.}\ \bibnamefont
  {Son}},\ }\bibfield  {title} {\bibinfo {title} {{Superconductivity by long
  range color magnetic interaction in high density quark matter}},\ }\href
  {https://doi.org/10.1103/PhysRevD.59.094019} {\bibfield  {journal} {\bibinfo
  {journal} {Phys. Rev. D}\ }\textbf {\bibinfo {volume} {59}},\ \bibinfo
  {pages} {094019} (\bibinfo {year} {1999})},\ \Eprint
  {https://arxiv.org/abs/hep-ph/9812287} {arXiv:hep-ph/9812287} \BibitemShut
  {NoStop}%
\bibitem [{\citenamefont {Evans}\ \emph
  {et~al.}(1999{\natexlab{a}})\citenamefont {Evans}, \citenamefont {Hsu},\ and\
  \citenamefont {Schwetz}}]{Evans:1998ek}%
  \BibitemOpen
  \bibfield  {author} {\bibinfo {author} {\bibfnamefont {N.~J.}\ \bibnamefont
  {Evans}}, \bibinfo {author} {\bibfnamefont {S.~D.~H.}\ \bibnamefont {Hsu}},\
  and\ \bibinfo {author} {\bibfnamefont {M.}~\bibnamefont {Schwetz}},\
  }\bibfield  {title} {\bibinfo {title} {{An Effective field theory approach to
  color superconductivity at high quark density}},\ }\href
  {https://doi.org/10.1016/S0550-3213(99)00175-3} {\bibfield  {journal}
  {\bibinfo  {journal} {Nucl. Phys. B}\ }\textbf {\bibinfo {volume} {551}},\
  \bibinfo {pages} {275} (\bibinfo {year} {1999}{\natexlab{a}})},\ \Eprint
  {https://arxiv.org/abs/hep-ph/9808444} {arXiv:hep-ph/9808444} \BibitemShut
  {NoStop}%
\bibitem [{\citenamefont {Evans}\ \emph
  {et~al.}(1999{\natexlab{b}})\citenamefont {Evans}, \citenamefont {Hsu},\ and\
  \citenamefont {Schwetz}}]{Evans:1998nf}%
  \BibitemOpen
  \bibfield  {author} {\bibinfo {author} {\bibfnamefont {N.~J.}\ \bibnamefont
  {Evans}}, \bibinfo {author} {\bibfnamefont {S.~D.~H.}\ \bibnamefont {Hsu}},\
  and\ \bibinfo {author} {\bibfnamefont {M.}~\bibnamefont {Schwetz}},\
  }\bibfield  {title} {\bibinfo {title} {{Nonperturbative couplings and color
  superconductivity}},\ }\href {https://doi.org/10.1016/S0370-2693(99)00093-3}
  {\bibfield  {journal} {\bibinfo  {journal} {Phys. Lett. B}\ }\textbf
  {\bibinfo {volume} {449}},\ \bibinfo {pages} {281} (\bibinfo {year}
  {1999}{\natexlab{b}})},\ \Eprint {https://arxiv.org/abs/hep-ph/9810514}
  {arXiv:hep-ph/9810514} \BibitemShut {NoStop}%
\bibitem [{\citenamefont {Sch\"afer}\ and\ \citenamefont
  {Wilczek}(1999)}]{Schafer:1998na}%
  \BibitemOpen
  \bibfield  {author} {\bibinfo {author} {\bibfnamefont {T.}~\bibnamefont
  {Sch\"afer}}\ and\ \bibinfo {author} {\bibfnamefont {F.}~\bibnamefont
  {Wilczek}},\ }\bibfield  {title} {\bibinfo {title} {{High density quark
  matter and the renormalization group in QCD with two and three flavors}},\
  }\href {https://doi.org/10.1016/S0370-2693(99)00162-8} {\bibfield  {journal}
  {\bibinfo  {journal} {Phys. Lett. B}\ }\textbf {\bibinfo {volume} {450}},\
  \bibinfo {pages} {325} (\bibinfo {year} {1999})},\ \Eprint
  {https://arxiv.org/abs/hep-ph/9810509} {arXiv:hep-ph/9810509} \BibitemShut
  {NoStop}%
\bibitem [{\citenamefont {{Shankar}}(2001)}]{2001JSP...103..485S}%
  \BibitemOpen
  \bibfield  {author} {\bibinfo {author} {\bibfnamefont {R.}~\bibnamefont
  {{Shankar}}},\ }\bibfield  {title} {\bibinfo {title} {{Luttinger
  Revisited{\textemdash}The Renormalization Group Approach}},\ }\href
  {https://doi.org/10.1023/A:1010333114332} {\bibfield  {journal} {\bibinfo
  {journal} {Journal of Statistical Physics}\ }\textbf {\bibinfo {volume}
  {103}},\ \bibinfo {pages} {485} (\bibinfo {year} {2001})},\ \Eprint
  {https://arxiv.org/abs/cond-mat/0009384} {arXiv:cond-mat/0009384
  [cond-mat.mes-hall]} \BibitemShut {NoStop}%
\bibitem [{\citenamefont {Kohn}(1959)}]{Kohn:1959zz}%
  \BibitemOpen
  \bibfield  {author} {\bibinfo {author} {\bibfnamefont {W.}~\bibnamefont
  {Kohn}},\ }\bibfield  {title} {\bibinfo {title} {{Image of the Fermi Surface
  in the Vibration Spectrum of a Metal}},\ }\href
  {https://doi.org/10.1103/PhysRevLett.2.393} {\bibfield  {journal} {\bibinfo
  {journal} {Phys. Rev. Lett.}\ }\textbf {\bibinfo {volume} {2}},\ \bibinfo
  {pages} {393} (\bibinfo {year} {1959})}\BibitemShut {NoStop}%
\bibitem [{\citenamefont {{Fetter}}\ and\ \citenamefont
  {{Walecka}}(1971)}]{1971qtmp.book.....F}%
  \BibitemOpen
  \bibfield  {author} {\bibinfo {author} {\bibfnamefont {A.~L.}\ \bibnamefont
  {{Fetter}}}\ and\ \bibinfo {author} {\bibfnamefont {J.~D.}\ \bibnamefont
  {{Walecka}}},\ }\href@noop {} {\emph {\bibinfo {title} {{Quantum theory of
  many-particle systems}}}}\ (\bibinfo {year} {1971})\BibitemShut {NoStop}%
\bibitem [{\citenamefont {Gor'kov}\ and\ \citenamefont
  {Melik-Barkhudarov}(1961)}]{Gorkov:1961}%
  \BibitemOpen
  \bibfield  {author} {\bibinfo {author} {\bibfnamefont {L.~P.}\ \bibnamefont
  {Gor'kov}}\ and\ \bibinfo {author} {\bibfnamefont {T.~K.}\ \bibnamefont
  {Melik-Barkhudarov}},\ }\bibfield  {title} {\bibinfo {title} {{Contribution
  to the Theory of Superfluidity in an Imperfect Fermi Gas}},\ }\href@noop {}
  {\bibfield  {journal} {\bibinfo  {journal} {Soviet Journal of Experimental
  and Theoretical Physics}\ }\textbf {\bibinfo {volume} {13}},\ \bibinfo
  {pages} {1018} (\bibinfo {year} {1961})}\BibitemShut {NoStop}%
\bibitem [{\citenamefont {Beane}\ \emph {et~al.}(2025)\citenamefont {Beane},
  \citenamefont {Capatti},\ and\ \citenamefont {Farrell}}]{Beane:2024tmd}%
  \BibitemOpen
  \bibfield  {author} {\bibinfo {author} {\bibfnamefont {S.~R.}\ \bibnamefont
  {Beane}}, \bibinfo {author} {\bibfnamefont {Z.}~\bibnamefont {Capatti}},\
  and\ \bibinfo {author} {\bibfnamefont {R.~C.}\ \bibnamefont {Farrell}},\
  }\bibfield  {title} {\bibinfo {title} {{Universal corrections to the
  superfluid gap in a cold Fermi gas}},\ }\href
  {https://doi.org/10.1103/PhysRevA.111.013309} {\bibfield  {journal} {\bibinfo
   {journal} {Phys. Rev. A}\ }\textbf {\bibinfo {volume} {111}},\ \bibinfo
  {pages} {013309} (\bibinfo {year} {2025})},\ \Eprint
  {https://arxiv.org/abs/2407.20168} {arXiv:2407.20168 [nucl-th]} \BibitemShut
  {NoStop}%
\bibitem [{\citenamefont {Chubukov}(1993)}]{PhysRevB.48.1097}%
  \BibitemOpen
  \bibfield  {author} {\bibinfo {author} {\bibfnamefont {A.~V.}\ \bibnamefont
  {Chubukov}},\ }\bibfield  {title} {\bibinfo {title} {Kohn-luttinger effect
  and the instability of a two-dimensional repulsive fermi liquid at t=0},\
  }\href {https://doi.org/10.1103/PhysRevB.48.1097} {\bibfield  {journal}
  {\bibinfo  {journal} {Phys. Rev. B}\ }\textbf {\bibinfo {volume} {48}},\
  \bibinfo {pages} {1097} (\bibinfo {year} {1993})}\BibitemShut {NoStop}%
\bibitem [{\citenamefont {Fujimoto}\ \emph {et~al.}(2020)\citenamefont
  {Fujimoto}, \citenamefont {Fukushima},\ and\ \citenamefont
  {Weise}}]{Fujimoto:2019sxg}%
  \BibitemOpen
  \bibfield  {author} {\bibinfo {author} {\bibfnamefont {Y.}~\bibnamefont
  {Fujimoto}}, \bibinfo {author} {\bibfnamefont {K.}~\bibnamefont
  {Fukushima}},\ and\ \bibinfo {author} {\bibfnamefont {W.}~\bibnamefont
  {Weise}},\ }\bibfield  {title} {\bibinfo {title} {{Continuity from neutron
  matter to two-flavor quark matter with $^1 S_0$ and $^3 P_2$
  superfluidity}},\ }\href {https://doi.org/10.1103/PhysRevD.101.094009}
  {\bibfield  {journal} {\bibinfo  {journal} {Phys. Rev. D}\ }\textbf {\bibinfo
  {volume} {101}},\ \bibinfo {pages} {094009} (\bibinfo {year} {2020})},\
  \Eprint {https://arxiv.org/abs/1908.09360} {arXiv:1908.09360 [hep-ph]}
  \BibitemShut {NoStop}%
\end{thebibliography}%

\end{document}